\documentclass[journal,comsoc]{IEEEtran}

\usepackage{graphicx,psfrag,epsfig,epsf,latexsym,hhline,amsmath,amssymb,multirow}
\usepackage[usenames,dvipsnames]{pstricks}
\usepackage{pst-plot}
\usepackage{pstricks-add}
\usepackage{pifont}
\usepackage{graphicx}
\usepackage{amsthm}
\usepackage{algorithm}
\usepackage{algpseudocode}
\usepackage[noadjust]{cite}
\usepackage{array}
\usepackage{blindtext}
\usepackage{etoolbox}
\usepackage{comment}
\usepackage{epstopdf}
\newcolumntype{M}[1]{>{\centering\arraybackslash}m{#1}}
\newcolumntype{P}[1]{>{\centering\arraybackslash}p{#1}}

\newcommand{\mc}[1]{\mathcal{#1}}

\newcommand{\mbb}[1]{\mathbb{#1}}

\newcommand{\msf}[1]{\mathsf{#1}}

\newcommand{\SNR}{\msf{SNR}}

\newcommand{\defeq}{\triangleq}

\newcommand{\E}{\mathbb{E}}

\newcommand{\iid}{i.\@i.\@d.\ }

\theoremstyle{definition}\newtheorem{lemma}{Lemma}
\theoremstyle{definition}\newtheorem{proposition}[lemma]{Proposition}
\theoremstyle{definition}
\theoremstyle{definition}

\newtheorem{example}[lemma]{Example}

\newtheorem{remark}[lemma]{Remark}

\begin{document}
\title{Downlink Non-Orthogonal Multiple Access without SIC for Block Fading Channels: An Algebraic Rotation Approach}
\author{Min Qiu,~\IEEEmembership{Student Member,~IEEE,}
Yu-Chih Huang,~\IEEEmembership{Member,~IEEE,}
and Jinhong Yuan,~\IEEEmembership{Fellow,~IEEE}


\thanks{This paper was presented in part at the IEEE International Conference on Communication 2019 \cite{MinICC19}.

M. Qiu, and J. Yuan are with the School of Electrical Engineering and Telecommunications, University of New South Wales, Sydney, NSW, 2052 Australia (e-mail: min.qiu@unsw.edu.au; j.yuan@unsw.edu.au).

Y.-C. Huang is with the Department of Communication Engineering, National Taipei University, New Taipei City 23741, Taiwan (e-mail: ychuang@mail.ntpu.edu.tw).
}%
}

\maketitle

\begin{abstract}
In this paper, we investigate the problem of downlink non-orthogonal multiple access (NOMA) over block fading channels. For the single antenna case, we propose a class of NOMA schemes where all the users' signals are mapped into $n$-dimensional constellations corresponding to the same algebraic lattices from a number field, allowing every user attains full diversity gain with single-user decoding, i.e., no successive interference cancellation (SIC). The minimum product distances of the proposed scheme with arbitrary power allocation factor are analyzed and their upper bounds are derived. Within the proposed class of schemes, we also identify a special family of NOMA schemes based on lattice partitions of the underlying ideal lattices, whose minimum product distances can be easily controlled. Our analysis shows that among the proposed schemes, the lattice-partition-based schemes achieve the largest minimum product distances of the superimposed constellations, which are closely related to the symbol error rates for receivers with single-user decoding. Simulation results are presented to verify our analysis and to show the effectiveness of the proposed schemes as compared to benchmark NOMA schemes. Extensions of our design to the multi-antenna case are also considered where similar analysis and results are presented.
\end{abstract}


\section{Introduction}
The fifth-generation (5G) networks are expected to support the explosive growth of smart devices along with the increasing demands of network access. In particular, they are required to offer higher data rate, spectral efficiency, and/or massive connectivity. One of the promising multiple access technique to meet these future requirements is non-orthogonal multiple access (NOMA). By allowing multiple users to share the same resource blocks and employing advanced multiuser detection technique at the receiver, NOMA is expected to provide enhanced system throughput and better user fairness than the current orthogonal multiple access (OMA) \cite{Dai15,Ding17J}.

According to the literature \cite{8114722,7676258,8085125}, NOMA can be categorized into code-domain and power-domain schemes. In this work, we focus on designing power-domain NOMA schemes for downlink multiuser transmission. For this area, extensive research has been conducted on designing user pairing \cite{ding17pair}, scheduling algorithms \cite{di16,Hsu2018VTC}, power allocation strategies \cite{8345745,Wei17,8529214}, the applications in multiple-input multiple-output (MIMO) systems \cite{7236924} and physical layer security \cite{7812773}. However, most of these works are based on Gaussian signaling which could arguably be infeasible for current wireless systems. The design of practical power-domain NOMA based on discrete and finite inputs has been considered in \cite{Choi2016,Shieh16,Fang16,Dong17,8291591,8517129}. Most notably, \cite{Fang16} develops lattice partition multiple access (LPMA) where the underlying codebook is constructed via Construction $\pi_A$ \cite{7962201} with a two-dimensional lattice partition corresponding to a Cartesian product of prime fields. Moreover, \cite{Shieh16,8291591} propose systematic designs of a $K$-user NOMA framework without SIC based on lattice partitions with one dimension and $n$ dimensions, respectively. While these schemes all rely on instantaneous channel state information (CSI) at the transmitter, we have considered a different scenario where only statistical CSI is available on the transmitter in \cite{8517129} and design coding and modulation schemes for $K$-user downlink NOMA for slow fading channels where the channel coherence time is significantly larger than the delay requirement.

In this work, we consider the task of designing reliable downlink NOMA schemes over block fading channels where the channel coherence time is larger than one symbol period but smaller than each packet transmission duration. This is a more realistic scenario compared to previous works, where the users within an urban macro cell are with low-mobility \cite{Huang:2011:MCC:2161741} or the cellular networks are using slow time-frequency hopping \cite{1564429}. For such a channel, an important performance metric is the diversity since it measures the decay rate in error probability with respect to the SNR. Our previous designs \cite{8291591,8517129}, however, achieve no diversity gains in block fading channels. Therefore, new approaches are required to design an efficient NOMA scheme to exploit the time diversity offered by the channel and to minimize the error probability. It is known that properly rotated QAM constellation can achieve full diversity order for point-to-point communication over block fading channels \cite{485720}. For multiuser transmission, the rotation angle of each user's constellation and the power allocation need to be carefully designed. Although there is some theoretical analysis on the ergodic rate achieved by random lattice ensembles \cite{7942023} with infinite dimensions, much is still lacking when it comes to the error probability and diversity order analysis of practical NOMA schemes based on discrete and finite inputs for this realistic scenario.

Very recently, the idea of rotation has been adopted in \cite{7880967} to design a two-user downlink NOMA system over block fading channels. In particular, the rotation angle of a user's QAM constellation is optimized through exhaustive search. However, the simulation results therein show that only the user whose constellation is optimized can obtain the diversity gain. Moreover, the diversity order of the system is 2, which is limited by the dimension of the underlying constellation. In addition, when the modulation order and its dimension become large, exhaustive search will become computationally prohibitive and require long processing time.

In this paper, the problem of achieving full diversity order and better error probability for every user in the downlink multiuser transmission over block fading channels is addressed. The main contributions of this paper are summarized as follows.
\begin{itemize}
\item We propose a class of downlink NOMA schemes without SIC for block fading channels with only statistical CSI at the transmitter and full CSI at the receiver. Specifically, the proposed scheme constructs an $n$-dimensional ideal lattice from algebraic number fields and carves its coset leaders to form the constellation for each user. This class of schemes is the first attempt to use algebraic methods to provide high reliability solutions to downlink multiuser communications. Within the proposed class, we also identify a special family of schemes that are closely related to lattice partitions of the base ideal lattice.

\item To evaluate the error performance of the proposed scheme under single-user decoding, we analyze the minimum product distance of the composite constellation of the proposed scheme. We first show the equivalence between the superimposed $n$-dimensional constellation carved from any ideal lattice and the Cartesian product of $n$ identical rotated superimposed one-dimensional constellation. As a result, we then rigorously prove that the minimum product distance of the $n$-dimensional composite constellation can be upper bounded by the minimum product distance of the equivalent one-dimensional superimposed constellation and derive the analytical expression for the upper bound as a function of all users' power allocation factors and spectral efficiencies. Moreover, our bound closely captures the actual minimum product distance in the sense that all the local maximums of the actual distance coincide with our upper bound. Furthermore, it is shown numerically that the special family of schemes corresponding to lattice partition can achieve the maximal minimum product distance among all the proposed schemes.

\item We then extend our analysis to the MIMO-NOMA system with orthogonal space-time block codes (OSTBC). For such codes, the probability of error is largely determined by the minimum determinant, which can be further simplified as a function of the minimum Euclidean distance of the underlying composite constellation. Following similar steps in our analysis in the single antenna case, we obtain the exact analytical expression of the minimum determinant of the superposition coded space-time codeword with arbitrary power allocation factors and spectral efficiencies. Again, a special family of schemes corresponding to lattice partition is identified and it achieves the maximal minimum determinant.

\item Simulation results are provided to illustrate that our scheme can provide a systematic design that each user employs the same ideal lattice and same rotation is sufficient to attain full diversity with single-user decoding (i.e., without SIC). Moreover, the special family of schemes based on lattice partitions provides substantially better error performance than the benchmark NOMA schemes.
\end{itemize}

\subsection{Notations in this paper}
Integers, real numbers, complex number and rational numbers are denoted by $\mathbb{Z}$, $\mathbb{R}$, $\mathbb{C}$ and $\mathbb{Q}$, respectively. Positive integers are represented by $\mathbb{Z}^+$. For $x \in \mathbb{R}$, $\left\lfloor x\right\rfloor$ rounds $x$ to the nearest integer that is not larger than $x$.

\section{System Model}\label{system_model}
In this work, we consider a downlink NOMA system where a base station wishes to broadcast $K$ messages $\mathbf{u}_1,\ldots,\mathbf{u}_K$ to $K$ users, one for each user. For $k\in\{1,\ldots,K\}$, the message $\mathbf{u}_k$ is a binary sequence of length $nm_k$, where $n$ is the dimension of the code and $m_k$ is the spectral efficiency of user $k$ in bits/s/Hz/real dimension. We emphasize here that due to the delay requirements, the channel between the transmitter and each user experiences independent block fading with a finite number of realizations within each data packet transmission duration, which is different from the slow fading model considered in \cite{8517129} where each user only gets to experience one realization within each data packet transmission duration.
For now, we assume that every device in the network is equipped with a single antenna and works in a half-duplex mode.

The base station encodes all users' messages $\mathbf{u}_1,\ldots,\mathbf{u}_K$ into a codeword $\mathbf{x} = [x[1],\ldots,x[n]]  \in \mathcal{M}$ of the codebook $\mathcal{M}\subset\mathbb{R}^n$, satisfying the power constraint $\E[\| \mathbf{x} \|^2] \leq n$. We denote by $\mathbf{h}_k = [h_k[1],\ldots,h_k[n]]\in\mbb{R}^n$ the instantaneous channel coefficient vector from the base station to user $k$. Here, each fading coefficient $h_k[l]$ is drawn i.i.d. from Rayleigh distribution. 
The received signal at user $k$ is denoted by $\mathbf{y}_k=[y_k[1],\ldots,y_k[n]]$ with
\begin{align}\label{eq:sm2}
y_k[l] = \sqrt{P}h_k[l]x[l]+z_k[l], \;  l= 1,\ldots, n,
\end{align}
where $P$ is the total power constraint at the base station and $z_k[l]\sim\mc{N}(0,1)$ is the Gaussian noise experienced at user $k$. Each user $k$ is assumed to have full CSI, i.e., $\mathbf{h}_k$, while the transmitter only has the statistical CSI, i.e., the distributions of each $\mathbf{h}_k$.
We note that this channel model is quite standard and can be easily obtained by \emph{interleaving} the codeword across multiple channel coherence time periods and applying de-interleaving and \emph{coherent detection} to the received signals \cite[Ch. 3.2]{tse_book}.

We measure the \emph{reliability} by the pairwise error probability (PEP). 
Following \cite[Ch. 3.2]{tse_book}, for any two codewords $\mathbf{x}_s,\neq \mathbf{x}_w \in \mathcal{M}$, user $k$'s error probability without SIC is upper bounded by the average PEP of the composite constellation over all $(\mathbf{x}_s,\mathbf{x}_w)$ pairs, which is 
\begin{align}\label{eq:pepZ}
P_e^{(k)} &\leq \frac{1}{|\mathcal{M}|}\sum_{s \neq w}\text{Pr}\left(\mathbf{x}_s \rightarrow \mathbf{x}_w|\mathbf{h}_k\right) \nonumber \\
&\leq \frac{1}{|\mathcal{M}|}\sum_{s \neq w}\prod_{l=1}^n \frac{1}{1+\overline{\SNR}_k(x_s[l]-x_w[l])^2/4} \nonumber \\
&< \frac{1}{|\mathcal{M}|}\sum_{s \neq w}\frac{4^{L_{(s,w)}}}{d_p(\mathbf{x}_s,\mathbf{x}_w)^2\overline{\SNR}_k^{L_{(s,w)}}} \nonumber \\ 
&\leq \frac{4^{\delta_L}(|\mathcal{M}|-1)}{\overline{\SNR}_k^{\delta_L}\min\limits_{s \neq w}\left\{d_p(\mathbf{x}_s,\mathbf{x}_w)^2 \right\}},
\end{align}
where
$d_p(\mathbf{x}_s,\mathbf{x}_w) \triangleq \prod_{s \neq w}|x_s[l]-x_w[l]|$
is the \emph{product distance} of $\mathbf{x}_s$ from $\mathbf{x}_w$ that differs in $L_{(s,w)} \leq n$ components and $\overline{\SNR_k} \triangleq \E[\|\mathbf{h}_k\|^2]P$ is the average SNR. It can be seen that in the high SNR regime, the overall error probability decreases exponentially with the order of $\delta_L \triangleq \min\limits_{s \neq w}\left\{L_{(s,w)}\right\}$, which is known as the \emph{diversity order}. The code has \emph{full diversity} when $\delta_L = n$. Moreover, one would like to maximize the minimum product distance $\min\limits_{s \neq w}\{d_p(\mathbf{x}_s,\mathbf{x}_w)\}$ in a bid to minimize the overall PEP, which provides additional \emph{coding gain} on top of the diversity gain. The diversity order and the minimum product distance are important metrics for improving the reliability of communication through block fading channels. Note that although we focus solely on Rayleigh fading channels in this paper, the diversity order and product distance criterion are generalizable to other fading channels, e.g., Rician fading \cite[Ch. 3.2]{tse_book}, \cite{Vucetic:2003:SC:861866,1256737}.

In this paper, we focus on the case of $K=2$ only as it is more practical for multi-carrier NOMA where each subcarrier is allocated to two users \cite{7273963,8449119}. This is also a common assumption in many works in the NOMA literature, see for example \cite{Wei17,8345745,Choi2016,Dong17}. We would like to emphasize that the schemes proposed in this paper are not limited to the two-user case and can be generalized to the general $K$-user case in a straightforward manner. However, the analysis becomes quite messy for $K>2$ and is thus left for future study. Throughout the paper, without loss of generality, we also assume that $\overline{\SNR_1}\geq \overline{\SNR_2}$ and thus users 1 and 2 are commonly referred to as the strong and weak users, respectively.

\section{Background}\label{sec:ANT}
In this section, we review some important concepts of algebraic number theory \cite{Oggier:2004:ANT:1166377.1166378} that will be useful later. The background on lattices can be found in \cite{conway1999sphere,Zamir15}.

An \emph{algebraic number} is a root of a monic polynomial (whose leading coefficient is 1) with coefficients in $\mathbb{Q}$. An \emph{algebraic integer} is a complex number which is a root of a monic polynomial with coefficients in $\mathbb{Z}$. A \emph{number field} $\mathbb{K} = \mathbb{Q}(\theta)$ is a field extension of $\mathbb{Q}$, where $\theta$ is an algebraic number and also a \emph{primitive element}. If this number field has degree $n$, then $\{1,\theta,\ldots,\theta^{n-1}\}$ is a \emph{basis} for $\mathbb{K}$. For this number field, there are $n$ distinct $\mathbb{Q}$-homomorphisms $\sigma_i: \mathbb{K} \rightarrow \mathbb{C}$ which is also called the \emph{embedding} of $\mathbb{K}$ into $\mathbb{C}$. The \emph{signature} of $\mathbb{K}$ is denoted by $(r_1,r_2)$ if among those $n = r_1+2r_2$ $\mathbb{Q}$-homomorphisms, there are $r_1$ real $\mathbb{Q}$-homomorphisms, i.e., $\sigma_1,\ldots,\sigma_{r_1}$, and $r_2$ pairs of complex $\mathbb{Q}$-homomorphisms, i.e., $\sigma_{r_1},\ldots,\sigma_n$, where $\sigma_{r_1+r_2+i}$ is the conjugate of $\sigma_{r_1+i}$ for $i \in \{1,\ldots,r_2\}$. Now for any $\varsigma = a_0 +a_1 \theta + \ldots+ a_{n-1} \theta^{n-1} \in \mathbb{K}$, the embedding of $\varsigma$ into $\mathbb{C}$ is given by
$\sigma_j(\varsigma) = \sigma_j\left(\sum\nolimits_{i=0}^{n-1} a_i \theta^i \right) = \sum\nolimits_{i=0}^{n-1} \sigma_j(a_i) \sigma_j(\theta)^i$ for $j\in \{1,\ldots,n \}$.
The algebraic norm of $\varsigma$ is given by $N(\varsigma) = \prod_{i=1}^n \sigma_i(\varsigma)$. The \emph{canonical embedding} $\Psi:\mathbb{K} \rightarrow \mathbb{R}^{r_1} \times \mathbb{C}^{r_2} \cong \mathbb{R}^n$ is defined by
$\Psi(\varsigma) = [\sigma_1(\varsigma),\ldots,\sigma_{r_1}(\varsigma),\sigma_{r_1+1}(\varsigma),\ldots,\sigma_{r_1+r_2}(\varsigma)]$, which is a ring homomorphism. 

Let $\mathcal{O}_\mathbb{K}$ be the ring of integers of $\mathbb{K}$, i.e., the set of all algebraic integers in $\mathbb{K}$. We define the \emph{discriminant} of $\mathbb{K}$ as
$d_\mathbb{K} \triangleq \det[(\sigma_j(\omega_i))_{i,j=1}^n]^2,$
where $\{\omega_1,\ldots,\omega_n \}$ is an integral basis of $\mathcal{O}_\mathbb{K}$. An \emph{algebraic lattice} $\Lambda  = \Psi(\mathcal{O}_\mathbb{K})$ is a lattice in $\mathbb{R}^{r_1} \times \mathbb{C}^{r_2} \cong \mathbb{R}^n$ with a generator matrix $\mathbf{G}_{\Lambda} = [(\Psi(\omega_i))_{i=1}^n ]$. A number field is said to be \emph{totally real} if it has signature $(r_1,r_2) = (n,0)$.
For a totally real number field $\mbb{K}$ of degree $n$ and an ideal $\mathcal{I} \subseteq \mathcal{O}_\mathbb{K}$ with an integral basis $\{\beta_1,\ldots, \beta_n\}$, the corresponding ideal lattice is given by $\Lambda = \Psi(\mathcal{I})$ which has the generator matrix $\mathbf{G}_{\Lambda} = [(\Psi(\beta_i))_{i=1}^n] \cdot \text{diag}(\sqrt{\sigma_1(\varsigma)},\ldots,\sqrt{\sigma_n(\varsigma)})$. It is shown in \cite{485720} that codes carved from an ideal lattice of a totally real number field attains the full diversity order $n$. Moreover, the minimum product distance of codes thus constructed can be easily guaranteed by the norm of the ideal $\mathcal{I}$.

\section{Downlink NOMA over block fading channels}\label{sec:proposed}
In this section, we first introduce the proposed class of NOMA schemes based on superpositions of codes from $n$-dimensional ideal lattices. We then identify, within the proposed class of schemes, a special family of schemes corresponding to lattice partitions of the underlying ideal lattices. The minimum product distance of the proposed schemes will be analyzed in Section~\ref{sec:perform_ana}.

\subsection{Proposed downlink NOMA schemes from ideal lattices}\label{A1}
Encouraged by the success of using ideal lattices for point-to-point communications over block-fading (see Section~\ref{sec:ANT}), we construct rotated version of multi-dimensional QAM (corresponding to $\mbb{Z}^n$ lattices) from a totally real ideal lattice. It is worth noting that the rotated versions of many other well-known lattices such as $D_4$, $E_6$, $E_8$ and $K_{12}$ that are good for block fading channels can also be constructed. Our choice of using rotated $\mbb{Z}^n$ is mainly for encoding/decoding complexity and for achieving full diversity order.

Throughout the paper, we use the cyclotomic construction \cite[Ch. 7.2]{Oggier:2004:ANT:1166377.1166378} to construct ideal lattices that are equivalent to $\mathbb{Z}^n$. Consider $\zeta = e^{\frac{2\pi \sqrt{-1}}{p}}$ the $p$-th primitive root of unity for some prime number $p \geq 5$. Construct $\mbb{K}=\mbb{Q}(\zeta + \zeta^{-1})$ the maximal real sub-field of the $p$-th cyclotomic field $\mbb{Q}(\zeta)$. This $\mbb{K}$ is totally real and has degree $n = \frac{p-1}{2}$. A set of integral basis is given by $\{\zeta+\zeta^{-1},\ldots, \zeta^n+\zeta^{-n} \}$. The $n$ embeddings of $\mbb{K}$ into $\mbb{C}$ are given by
\begin{align}
    \sigma_j(\zeta^i+\zeta^{-i}) = \zeta^{ij}+\zeta^{-ij} = 2\cos\left( \frac{2\pi i j}{p}\right), \; i,j \in \{1,\ldots,n\}.
\end{align}
Then, the generator matrix is given by
\begin{align}\label{G_ideal}
\mathbf{G}_{\Lambda} = \frac{1}{\sqrt{p}}\mathbf{T}\cdot [(\sigma_j(\zeta^i+\zeta^{-i}))_{i,j=1}^n] \cdot \text{diag}(\sqrt{\sigma_1(\varsigma)},\ldots,\sqrt{\sigma_n(\varsigma)}),
\end{align}
where $\mathbf{T}$ is an upper triangular matrix with entries $t_{i,j}=1$ for $i \leq j$; $\varsigma = (1-\zeta)(1-\zeta^{-1})$ is to ensure that $\Lambda$ is equivalent to $\mathbb{Z}^n$; and $\frac{1}{\sqrt{p}}$ is to normalize the volume of $\Lambda$ such that $\text{Vol}(\Lambda)=1$.

The minimum product distance of this family of ideal lattices is
\begin{align}\label{eq:dpcal}
d_{p,\min}(\Lambda) = \sqrt{\frac{\det(\mathbf{G}_{\Lambda})}{d_\mathbb{K}}} = p^{-\frac{n-1}{2}}.
\end{align}

Having constructed the considered ideal lattice, we now introduce the encoding and decoding steps of our proposed NOMA scheme at the transmitter and the receiver, respectively.
\subsubsection{Transmitter side}\label{sec:enc}
For user $k\in\{1,2\}$, a subset $\mc{C}_k$ of the ideal lattice is carved to form the constellation of the user $k$. Specifically, $\mc{C}_k$ has cardinality $2^{nm_k}$ and is the complete set of coset leaders (see \cite[Example 2.3.1]{Zamir15} for the definition) of the lattice partition $\Lambda/2^{m_k}\Lambda$. Here, $m_k$ is the target spectral efficiency for user $k$ in bits/s/Hz/real dimension. User $k$'s message $\mathbf{u}_k$ is mapped into $\mathbf{v}_k\in\mc{C}_k$. 
The transmitter then sends the superimposed signal $\mathbf{x}=\eta(\sqrt{\alpha}\mathbf{v}_1+\sqrt{1-\alpha}\mathbf{v}_2-\mathbf{d})$, where
\begin{align}\label{eq:sup1}
\mathbf{x} \in \eta(\mathcal{C}-\mathbf{d} )&= \eta(\sqrt{\alpha}\mathcal{C}_1+\sqrt{1-\alpha}\mathcal{C}_2-\mathbf{d}) \nonumber \\
&= \eta(\sqrt{\alpha}(\mathcal{C}_1-\mathbf{d}_1)+\sqrt{1-\alpha}(\mathcal{C}_2-\mathbf{d}_2)),
\end{align}
where $\mathbf{d}_1= \E[\mathcal{C}_1]$, $\mathbf{d}_2= \E[\mathcal{C}_2]$, and $\mathbf{d} = \E[\sqrt{\alpha}\mathcal{C}_1+\sqrt{1-\alpha}\mathcal{C}_2]$ are length $n$ dither vectors to ensure the constellations $\mc{C}_1$, $\mc{C}_2$, and $\mc{C}$, respectively, to have zero mean; $\eta$ is a normalize factor for ensuring power constraint $\mbb{E}[\|\mathbf{x}\|^2]\leq n$;
and $\alpha,1-\alpha \in[0,1]$ are the power allocation factors for users 1 and 2, respectively. Here, the normalization factor $\eta$ is computed by using Lemma \ref{lem:zn_P} in Appendix \ref{appendix:lemma} as
\begin{align}\label{normalize2}
\eta =& \sqrt{\frac{n}{\E[\|\mathcal{C}-\mathbf{d} \|^2]}} \nonumber \\
 \overset{\eqref{eq:sup1}}=& \sqrt{\frac{n}{\E[\|\sqrt{\alpha}(\mathcal{C}_1-\mathbf{d}_1)+\sqrt{1-\alpha}(\mathcal{C}_2-\mathbf{d}_2) \|^2]}} \nonumber \\
 =& \sqrt{\frac{n}{\alpha\E[\|\mathcal{C}_1-\mathbf{d}_1\|^2]+(1-\alpha)\E\|\mathcal{C}_2-\mathbf{d}_2 \|^2]}} \nonumber \\
 \overset{\eqref{eq:zn_P}}=& \sqrt{\frac{n}{\alpha\frac{n}{12}(2^{2m_1}-1)+(1-\alpha)\frac{n}{12}(2^{2m_2}-1)}} \nonumber \\
 =& \sqrt{\frac{12}{(2^{2m_1}-2^{2m_2})\alpha+2^{2m_2}-1}}.
\end{align}

\subsubsection{Receiver side}
Recall that the received message at user $k\in \{1,2\}$ is denoted by $\mathbf{y}_k$ and is given in \eqref{eq:sm2}. There are two options for the decoder, depending on the implementation and application. 
If a single-user decoder is adopted (i.e., without performing SIC), the decoder of user $k$ attempts to recover $\mathbf{u}_k$ from $\mathbf{y}_k$ by treating the other user's signal as interference. If an SIC decoder is adopted, user 2 remains the same decoding procedure, while user 1 first decodes $\mathbf{u}_2$, subtracts it out, and then decodes its own message. Both single-user decoding and SIC decoding
will be included in simulations for comparison. However, our design and analysis focus solely on the case with single-user decoding as it is one of the main motivation of this work.

\begin{remark}
Similar to most works considering block Rayleigh fading channels (see \cite{Oggier:2004:ANT:1166377.1166378} and reference therein), we focus solely on diversity order and minimum product distance. It is worth mentioning that standard channel coding can be employed on top of the modulation schemes of this work to obtain additional coding gain at the cost of further lowering the spectral efficiency.
\end{remark}

\begin{remark}
Consider a $K$-user downlink NOMA system. For the conventional power-domain NOMA, each user would have to decode other $(K-1)$ users' messages to perform SIC because each user has some probability to potentially become the strongest channel user. Thus, the demodulation and decoding delay can be as large as $K$ times of that for our proposed scheme without SIC. Moreover, encoding delays are introduced by SIC as a result of re-encoding the decoded message and then re-mapping the codeword to the modulation. In contrast, re-encoding and re-mapping are not required in our scheme without SIC.
\end{remark}

\subsection{Proposed schemes based on lattice partitions}\label{Sec:LP}
Now, we identify a special family of the proposed schemes within the proposed class of schemes. In this family of schemes, after the mapping process from $\mathbf{u}_k$ to $\mathbf{v}_k\in\mathcal{C}_k$ for $k\in\{1,2\}$, the transmitted signal is given by
\begin{align}\label{eqn:x_def}
\mathbf{x}' = \eta'\left(\mathbf{v}_1+2^{m_1}\mathbf{v}_2-\mathbf{d}' \right)\in\eta'\left( \mc{C}_1 + 2^{m_1}\mc{C}_2-\mathbf{d}' \right),
\end{align}
where $\mathbf{d}'$ is a deterministic dither to ensure the composite constellation $\mc{C}'=\mc{C}_1 + 2^{m_1}\mc{C}_2$ have zero mean and
\begin{equation}\label{normalize_LP}
    \eta'=\sqrt{\frac{12}{2^{2(m_1+m_2)}-1}},
\end{equation}
is the normalization factor to ensure the power constraint $\mbb{E}[\|\mathbf{x}'\|^2]\leq n$. To see that $\eta'$ is indeed the correct normalization factor, we use Lemma \ref{lem:zn_P} in Appendix \ref{appendix:lemma} to obtain that $\mbb{E}[\|\mc{C}'-\mathbf{d}'\|^2]=\frac{n}{12}(2^{2(m_1+m_2)}-1)$.
Here, the power allocation is $\alpha =  \frac{1}{1+2^{2m_1}}$. When substituting this power allocation into (6) and decomposing $\mathbf{d}' = \sqrt{\alpha}\mathbf{d}_1+\sqrt{1-\alpha}\mathbf{d}_2$, it can be easily verified that this family of schemes described in \eqref{eqn:x_def} is a special case of the proposed class of schemes in \eqref{eq:sup1}.


The beauty of this family of schemes is that the composite constellation $\mc{C}'$ corresponds to the lattice partition $\Lambda/2^{m_1+m_2}\Lambda$ because $\Lambda/2^{m_1}\Lambda+2^{m_1}(\Lambda/2^{m_2}\Lambda) = \Lambda/2^{m_1+m_2}\Lambda$ for $\Lambda$ equivalent to $\mathbb{Z}^n$. Moreover, the relationship among $\mc{C}_1$, $\mc{C}_2$, and $\mc{C}'$ closely follows the lattice partition chain $\Lambda/2^{m_1}\Lambda/2^{m_1+m_2}\Lambda$ and hence many nice properties of the underlying ideal lattice $\Lambda$ naturally carry over to the individual and composite constellations. For example, since the superimposed constellation still preserves the nice lattice structure, efficient lattice decoders such as the sphere decoder \cite{771234} can be used at each receiver for decoding. Also, the minimum product distance of a scheme within this family can be precisely computed as shown in the following proposition.
\begin{proposition}\label{prop:LP}
The lattice-partition scheme with ideal lattices as the base lattice can provide full diversity \emph{to each user} and the composite constellation $\eta'(\mc{C}'-\mathbf{d}')$ has a minimum product distance
\begin{align}\label{dminp_LP111}
d_{p,\min}(\eta'(\mc{C}'-\mathbf{d}')) = \left(\frac{12}{2^{2(m_1+m_2)}-1}\right)^{\frac{n}{2}}d_{p,\min}(\Lambda).
\end{align}
\end{proposition}
\begin{IEEEproof}
Since $\mc{C}'$ corresponds to the lattice partition $\Lambda/2^{m_1+m_2}\Lambda$, the minimum product distance of the composite constellation can be derived as
\begin{align}\label{eq:dp_lattice}
    d_{p,\min}(\eta'(\mc{C}'-\mathbf{d}'))=&d_{p,\min}(\eta'\Lambda) 
    =(\eta')^n\sqrt{\frac{\det(\Lambda)}{d_\mathbb{K}}}\nonumber \\
    \overset{(a)}=&\left(\frac{12}{2^{2(m_1+m_2)}-1}\right)^{\frac{n}{2}}d_{p,\min}(\Lambda),
\end{align}
where $(a)$ is obtained by plugging $\eta'$ from \eqref{normalize_LP}.

Now, since $ d_{p,\min}(\eta'(\mc{C}'-\mathbf{d}'))>0$, full diversity is thus guaranteed according to \eqref{eq:pepZ}.
\end{IEEEproof}

\section{Performance analysis}\label{sec:perform_ana}
In this section, we analyze $d_{p,\min}(\eta(\mathcal{C}-\mathbf{d}))$, the minimum product distance of the \emph{normalized} and \emph{dithered} composite constellation $\eta(\mathcal{C}-\mathbf{d})$ defined in \eqref{eq:sup1} for any parameters $m_1$, $m_2$, $n$, $\alpha$. We emphasize that under block fading, the symbol error rate (SER) performance of the whole downlink system is closely related to $d_{p,\min}(\eta(\mathcal{C}-\mathbf{d}))$ according to \eqref{eq:pepZ}. Moreover, the analytical results of the minimum product distances will provide insights into the relationship between spectral efficiency, power allocation factor and the error performance of the proposed scheme. In what follows, we first introduce a few preparations and definitions in Sec.~\ref{sec:def}. We then present the main results of this section in Sec.~\ref{sec:Main_result}, followed by a rigorous proof in Sec. \ref{sec:dminp_ana}.

\subsection{Preparations and definitions}\label{sec:def}
\subsubsection{}\label{def:layer} 
\begin{figure}[ht!]
	\centering
\includegraphics[width=3.04in,clip,keepaspectratio]{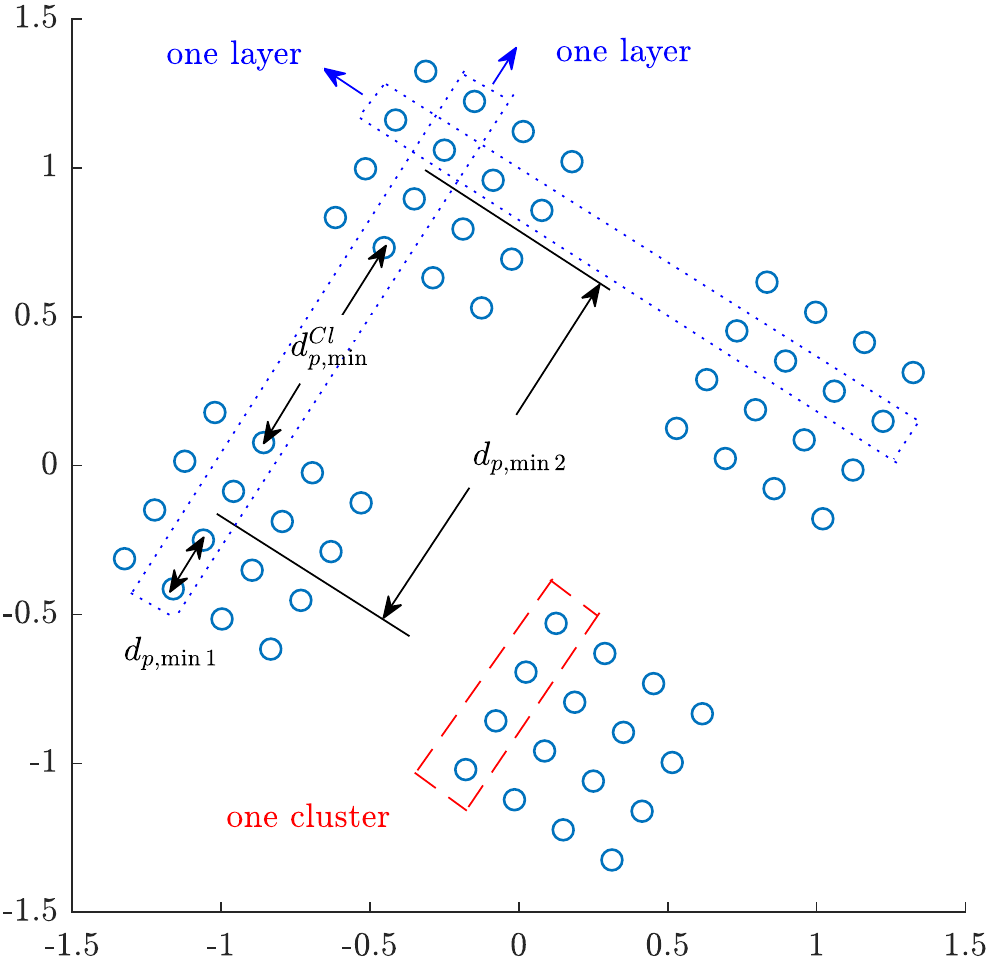}
\caption{An example of a superimposed constellation with $\Lambda$ being a two-dimensional ideal lattice and $(m_1,m_2) = (2,1)$.}
\label{fig:2}
\end{figure}

We define a \emph{layer} of $\eta(\mc{C}-\mathbf{d})$ in \eqref{eq:sup1} to be the collection of points constituting a shifted version of a rotated and dithered one-dimensional superimposed constellation
\begin{align}\label{eq:1dsup}
\eta(\mathcal{X}-d^*)\mathbf{R} = \eta(\sqrt{\alpha}(\mathcal{X}_1-d^*_1)+\sqrt{1-\alpha}(\mathcal{X}_2-d^*_2))\mathbf{R},
\end{align}
where $\mathcal{X}_k$ is a complete set of the coset leaders of the one-dimensional lattice partition $\mathbb{Z}/2^{m_k}\mathbb{Z}$, $d^*_k = \E[\mathcal{X}_k]$ is a scalar dither for $k \in\{1,2\}$, and $\mathbf{R}$ is an $n \times n$ rotation matrix such that the shifted and rotated one-dimensional constellation becomes a subset of $\eta(\mc{C}-\mathbf{d})$. In other words, a layer is given by $\{[\lambda_1,\dots,\lambda_n]\mathbf{R}| \lambda_j \in \eta(\mathcal{X}-d^*)  \}$ for some fixed $\lambda_1,\ldots,\lambda_{j-1},\lambda_{j+1},\ldots,\lambda_n\in \eta(\mathcal{X}-d^*)$. Examples of all the layers for the case of $(m_1,m_2) = (2,1)$ and $n=2$ are illustrated in Fig. \ref{fig:2} where each circle represents a constellation point of $\eta(\mc{C}-\mathbf{d})$ and there are 16 layers in total.

\subsubsection{}\label{def:ilmpd} We denote by $d_{p,\min}(\eta(\mathcal{X}-d^*)\mathbf{R})$ the \emph{intra-layer minimum product distance} as the minimum product distance between any pair of two distinct constellation points within a layer, i.e., within the shifted version of the (rotated) one-dimensional constellation $\eta(\mathcal{X}-d^*)\mathbf{R}$.

\subsubsection{}\label{def:cl} We define a \emph{cluster} to be all the points in a shifted version of user 1's constellation \emph{in one layer}, i.e., $\text{Cl}_{\nu} \triangleq \{\sqrt{\alpha}(\mathcal{X}_1-d^*_1)\mathbf{R}+\nu\}$ for a fixed $\nu \in \sqrt{1-\alpha}(\mathcal{X}_2-d^*_2)\mathbf{R}$. Each layer has $2^{m_2}$ clusters. In the example shown in Fig. \ref{fig:2}, there are 2 clusters inside a layer. With a slight abuse of notation, we define the minimum product distance between two distinct clusters $\text{Cl}_{\nu}$ and $\text{Cl}_{\mu}$ as
\begin{equation}
    d_{p,\min}(\text{Cl}_{\nu},\text{Cl}_{\mu}) \defeq \min\limits_{ \boldsymbol{\lambda}_1 \in \text{Cl}_{\nu},\boldsymbol{\lambda}_2 \in \text{Cl}_{\mu}} \{d_p(\boldsymbol{\lambda}_1,\boldsymbol{\lambda}_2)\}.
\end{equation}

\subsubsection{}\label{def:icmpd} The \emph{inter-cluster minimum product distance} is then defined as
\begin{equation}\label{eq:ICPD}
  d_{p,\min}^{\text{Cl}} \triangleq \min\limits_{\nu,\mu \in \sqrt{1-\alpha}(\mathcal{X}_2-d^*_2)\mathbf{R}} \{d_{p,\min}(\text{Cl}_{\nu},\text{Cl}_{\mu}) \}.
\end{equation}
An example of $d_{p,\min}^{\text{Cl}}$ can also be found in Fig. \ref{fig:2}. 

\subsubsection{}\label{def:d2d2} We denote by $d_{p,\min 1}$ and $d_{p,\min 2}$ the minimum product distance of users 1 and 2's constellations \emph{in one layer}, respectively. To be specific, they are computed as
\begin{align}
d_{p,\min1}  \triangleq& d_{p,\min}(\eta\sqrt{\alpha}(\mathcal{X}_1-d^*_1)\mathbf{R})  \nonumber \\
=& (\eta\sqrt{\alpha})^n d_{p,\min}((\mathcal{X}_1-d^*_1)\mathbf{R}), \label{eq:dp1_def} \\
d_{p,\min2}  \triangleq& d_{p,\min}(\eta\sqrt{1-\alpha}(\mathcal{X}_1-d^*_1)\mathbf{R}) \nonumber \\
=&(\eta\sqrt{1-\alpha})^n d_{p,\min}(\mathcal{X}_1-d^*_1)\mathbf{R}). \label{eq:dp2_def}
\end{align}
Examples of $d_{p,\min1}$ and $d_{p,\min2}$ are shown in Fig. \ref{fig:2}.

\subsection{Main result of this section}\label{sec:Main_result}
The minimum product distance of the proposed NOMA scheme with arbitrary power allocation is upper bounded as follows.
\begin{proposition}\label{prop:AP}
The minimum product distance of the proposed NOMA scheme with ideal lattices as the base lattice and with arbitrary power allocation $\alpha\in[0,1]$ is upper bounded by
\begin{figure*}[t]
\begin{align}\label{eq:case2}
d_{p,\min}(\eta(\mathcal{C}-\mathbf{d})) \leq 
  \left\{\begin{array}{ll}
d_{p,\min 1}, &\alpha\in [0 ,\frac{1}{1+2^{2m_1}}]\\
d_{p,\min}^{\text{Cl}(m_2=1)},
&\alpha \in \big(\frac{1}{1+2^{2m_1}},\frac{4}{(2^{m_1}-\frac{1}{2})+4}\big], \\
\multirow{2}{*}{$\min\limits_{\substack{\gamma \in \{ 0,\ldots,\lfloor\frac{\xi-1}{2} \rfloor\}, \\ \beta \in \{1,\ldots,\xi-1\}}} \left\{\left|\gamma\sqrt[n]{d_{p,\min 1}} -\beta\sqrt[n]{d_{p,\min}^{\text{Cl}(m_2=1)}}\right|^n  \right\},$} 
 &\alpha \in \Big(\frac{(\xi-1)^2}{\left(2^{m_1}-\frac{1}{2} \right)^2+(\xi-1)^2}, 
\frac{\xi^2}{\left(2^{m_1}-\frac{1}{2} \right)^2+\xi^2}  \Big],\\
&\xi = 3, \ldots,2^{m_2}-1 \\
\min\limits_{\substack{\gamma \in \{ 0,\ldots,\frac{2^{m_2}}{2}-1 \}, \\ \beta \in \{1,\ldots,2^{m_2}-1\}}} \left\{\left|\gamma\sqrt[n]{d_{p,\min 1}} -\beta\sqrt[n]{d_{p,\min}^{\text{Cl}(m_2=1)}}\right|^n \right\},
& \alpha \in \big(\frac{(2^{m_2}-1)^2}{(2^{m_1}-\frac{1}{2})^2+(2^{m_2}-1)^2},\frac{1}{2}\big]  \\
\end{array} \right.
\end{align}
\hrule
\end{figure*}
where
\begin{align}
d_{p,\min1}   =& \left(\frac{12\alpha}{(2^{2m_1}-2^{2m_2})\alpha+2^{2m_2}-1}\right)^{\frac{n}{2}}d_{p,\min}(\Lambda), \label{eq:dp1} \\
d_{p,\min2}   =& \left(\frac{12(1-\alpha)}{(2^{2m_1}-2^{2m_2})\alpha+2^{2m_2}-1}\right)^{\frac{n}{2}}d_{p,\min}(\Lambda), \label{eq:dp2} 
\end{align}

\begin{figure*}[t]
\begin{align}\label{eq:case1}
d_{p,\min}^{\text{Cl}(m_2=1)} = \left\{\begin{array}{ll}
\left|\sqrt[n]{d_{p,\min 2}}- (2^{m_1}-1)\sqrt[n]{d_{p,\min 1}}\right|^n, 
&\alpha \in \big(\frac{1}{1+2^{2m_1}}, \frac{1}{(2^{m_1}-\frac{3}{2})^2+1}\big] \\
\multirow{2}{*}{$\left|\sqrt[n]{d_{p,\min 2}}- (2^{m_1}-l)\sqrt[n]{d_{p,\min 1}}\right|^n,$} &\alpha \in \big(\frac{1}{(2^{m_1}+\frac{1}{2}-l)^2+1},\frac{1}{(2^{m_1}-\frac{1}{2}-l)^2+1}\big],\\
&l = 2, \ldots,2^{m_1}-2  \\
\left(\sqrt[n]{d_{p,\min 2}} - \sqrt[n]{d_{p,\min 1}}\right)^n, 
&\alpha \in (\frac{4}{13},\frac{1}{2}] \\
\end{array} \right., 
\end{align}
\hrule
\end{figure*}
and $d_{p,\min}^{\text{Cl}(m_2=1)}$ denotes $d_{p,\min}^{\text{Cl}}$ the inter-cluster minimum product distance for the case of $m_2 = 1$ for any $m_1 \in \mathbb{Z}^+$. The the upper bound for $\alpha\in [\frac{1}{2},1]$ can be obtained by switching the roles of $m_1$ and $m_2$ and substituting $1-\alpha$ into $\alpha$ from \eqref{eq:case2}.
\end{proposition}

The proof of this proposition is described in details in Section \ref{sec:dminp_ana}. Before that, we would like to emphasize that one can find the exact minimum product distance of $\eta(\mathcal{C}-\mathbf{d})$ for a given $\alpha \in [0,1]$ by numerically calculating all the product distances between all pairs of two constellation points in $\eta(\mathcal{C}-\mathbf{d})$ and find the minimum value among them. However, the computational complexity will dramatically increase with $m_1$, $m_2$ and $n$ increasing. We use the following example to demonstrate the effectiveness of our analytical upper bound.

\begin{example}\label{exa:dpmin}
Consider $(m_1,m_2) = (3,3)$ and $n=2$. Both $\mc{C}_1-\mathbf{d}_1$ and $\mc{C}_2-\mathbf{d}_2$ are rotated 64-QAM constellations and $\mc{C}-\mathbf{d}$ becomes a superimposed constellation with 4096 constellation points. In Fig. \ref{fig:7}, we evaluate the upper bound of $d_{p,\min}(\eta(\mathcal{C}-\mathbf{d}))$ in \eqref{eq:case2} and the exact values of $d_{p,\min}(\eta(\mathcal{C}-\mathbf{d}))$ by computer search for $\alpha \in [0,0.5]$. The minimum product distance achieved by our scheme based on lattice partition is also plotted.
\begin{figure}[ht!]
	\centering
\includegraphics[width=3.4in,clip,keepaspectratio]{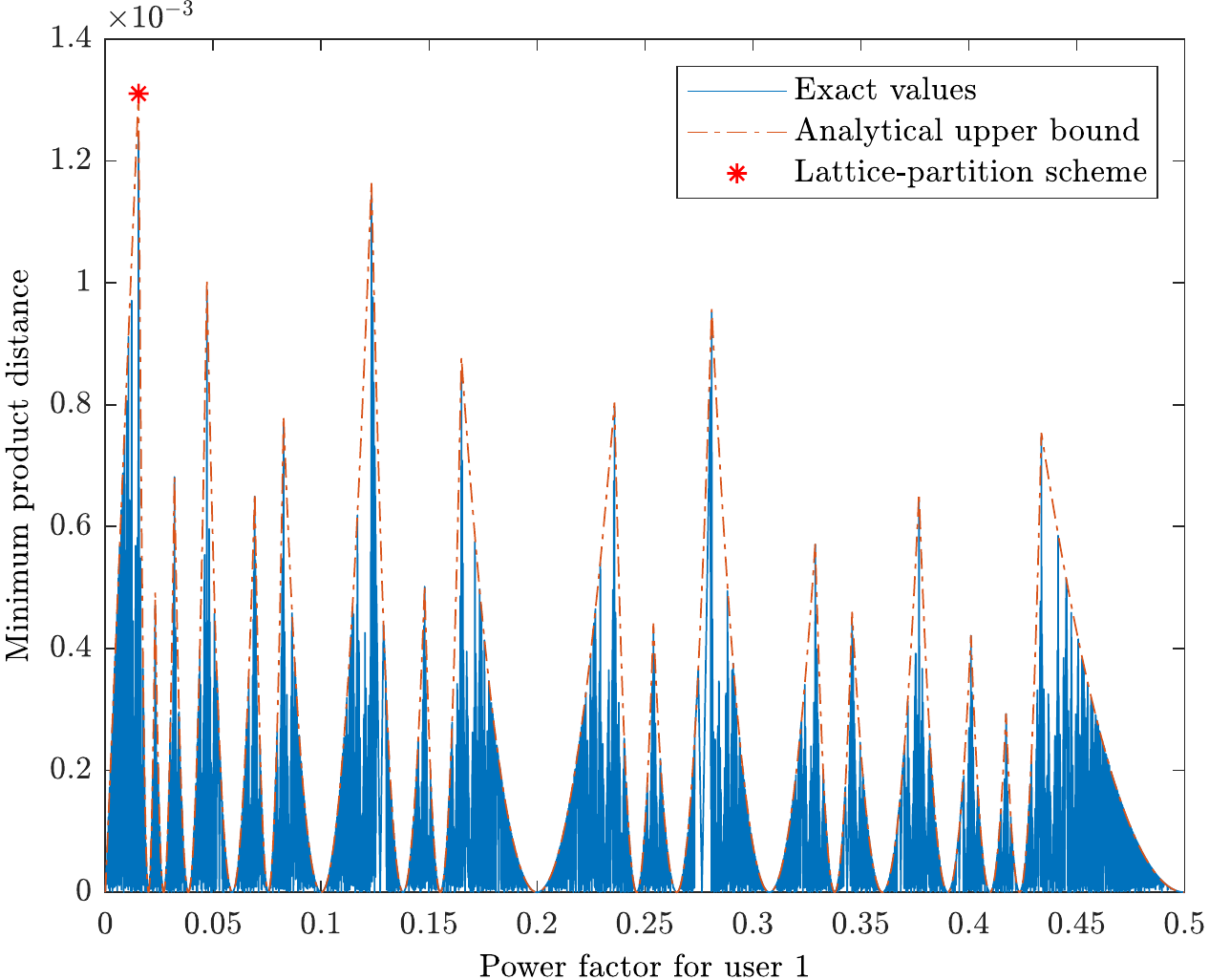}
\caption{Minimum product distances of the scheme considered in Example~\ref{exa:dpmin} with various $\alpha\in[0,0.5]$.}
\label{fig:7}
\end{figure}

It can be observed that the derived upper bound well captures the trend of the changes in $d_{p,\min}(\eta(\mathcal{C}-\mathbf{d}))$ and fits all the local maximum points (peak values in the figure). Most notably, the proposed scheme based on lattice partition achieves the largest value (the first peak value in the figure), which shows the optimality of this scheme. Although we have not rigorously proved that this scheme is always optimal for a general pair of $(m_1,m_2)$, it is optimal for all the cases that we have tested, including every $(m_1,m_2)$ for $m_1,m_2\in\{1,\ldots,8\}$ (each user's constellation ranging from rotated 4-QAM to rotated $2^{16}$-QAM).
\end{example}

\subsection{Proof of Proposition~\ref{prop:AP}}\label{sec:dminp_ana}
We now derive the upper bound for the minimum product distance of the superimposed constellation $d_{p,\min}(\eta(\mathcal{C}-\mathbf{d}))$. 
In what follows, we first prove in Lemma~\ref{the:1} that the $n$-dimensional superimposed constellation is an $n$-fold cartesian product of a one-dimensional \emph{superimposed} constellation. Then, we show in Lemma~\ref{the:2a} that $d_{p,\min}(\eta(\mathcal{C}-\mathbf{d}))$ can be upper bounded by the minimum product distance of this one-dimensional superimposed constellation. With these lemmas, we then bound the minimum product distance of $d_{p,\min}(\eta(\mathcal{C}-\mathbf{d}))$ by analyzing the minimum product distances of the one-dimensional superimposed constellation. 

\begin{lemma}\label{the:1}
Consider the constellation $\eta(\mathcal{C}-\mathbf{d})$ defined in \eqref{eq:sup1} for $\alpha \in [0,1]$ and the base lattice $\Lambda$ is equivalent to $\mathbb{Z}^n$. The constellation $\eta(\mathcal{C}-\mathbf{d})$ is the rotated $n$-fold Cartesian product of the one-dimensional constellation $\eta(\sqrt{\alpha}(\mathcal{X}_1-d^*_1)+\sqrt{1-\alpha}(\mathcal{X}_2-d^*_2))$ in \eqref{eq:1dsup}.
\end{lemma}

\begin{IEEEproof}
Following \eqref{eq:sup1}, we write the superimposed constellation as
\begin{align}
&\eta(\mathcal{C}-\mathbf{d}) = \eta(\sqrt{\alpha}(\mathcal{C}_1-\mathbf{d}_1)+\sqrt{1-\alpha}(\mathcal{C}_2-\mathbf{d}_2)) \nonumber \\
 \overset{(a)}= & \{\eta(\sqrt{\alpha}(\mathbf{b}_1\mathbf{G}_{\mathbb{Z}^n}-\mathbf{d}^*_1)\mathbf{R}+\sqrt{1-\alpha}(\mathbf{b}_2\mathbf{G}_{\mathbb{Z}^n}-\mathbf{d}^*_2)\mathbf{R} )\} \nonumber \\
 \overset{(b)}=& \{\eta(\sqrt{\alpha}(\mathbf{b}_1-\mathbf{d}^*_1)+\sqrt{1-\alpha}(\mathbf{b}_2-\mathbf{d}^*_2))\mathbf{R} \} \nonumber \\
 \overset{(c)} =& (\eta(\sqrt{\alpha}(\mathcal{X}_1-d^*_1[1])+\sqrt{1-\alpha}(\mathcal{X}_2-d^*_2[1])) \nonumber \\ &\times\eta(\sqrt{\alpha}(\mathcal{X}_1-d^*_1[2])+\sqrt{1-\alpha}(\mathcal{X}_2-d^*_2[2])) \nonumber \\
&\times \ldots \times  \eta(\sqrt{\alpha}(\mathcal{X}_1-d^*_1[n])+\sqrt{1-\alpha}(\mathcal{X}_2-d^*_2[n])))\mathbf{R}  \nonumber \\
 \overset{(d)} =& (\eta(\sqrt{\alpha}(\mathcal{X}_1-d^*_1)+\sqrt{1-\alpha}(\mathcal{X}_2-d^*_2)) \nonumber \\
 &\times \eta(\sqrt{\alpha}(\mathcal{X}_1-d^*_1)+\sqrt{1-\alpha}(\mathcal{X}_2-d^*_2)) \nonumber \\
&\times \ldots \times  \eta(\sqrt{\alpha}(\mathcal{X}_1-d^*_1)+\sqrt{1-\alpha}(\mathcal{X}_2-d^*_2)))\mathbf{R},
\end{align}
where $(a)$ follows that $\mathcal{C}_k-\mathbf{d}_k$ is obtained by multiplying the dithered coset leaders of $\mathbb{Z}^n/2^{m_k}\mathbb{Z}^n$ to the rotational matrix $\mathbf{R}$ while these coset leaders are generated by $(\mathbf{b}_k\mathbf{G}_{\mathbb{Z}^n} - \mathbf{d}^*_k)$ with $\mathbf{b}_k = [b_k[1],b_k[2],\ldots,b_k[n]] \in \mathbb{Z}^n$ and $\mathbf{d}^*_k = \E[\{\mathbf{b}_k\mathbf{G}_{\mathbb{Z}^n} \}]$ for $k = 1,2$; $(b)$ follows that since $\mathbf{G}_{\mathbb{Z}^n} = \mathbf{I}_n$, thus $\mathbf{b}_k \in \{\boldsymbol{\lambda} \; \text{mod} \; 2^{m_k}\mathbb{Z}^n, \boldsymbol{\lambda} \in \mathbb{Z}^n\}$ and $\mathbf{d}^*_k = \E[\{\mathbf{b}_k \}]$ for $k = 1,2$; $(c)$ is due to that each component of $\mathbf{b}_k$ follows $b_k[i] \in \mathcal{X}_k = \{\lambda \; \text{mod} \; 2^{m_k}\mathbb{Z}, \lambda \in \mathbb{Z}\}$ for $k =1,2$ and $i = 1,\ldots,n$ because $\mathbf{b}_k$, the coset leader of $\mathbb{Z}^n/2^{m_k}\mathbb{Z}^n$, is precisely the $n$-fold Cartesian product of the coset leader of $\mathbb{Z}/2^{m_k}\mathbb{Z}$; and $(d)$ follows that $d^*_k[1]= d^*_k[2]= \ldots = d^*_k[n] = d^*_k$ for $k = 1,2$ because
\begin{align}
&[d^*_k[1],d^*_k[2],\ldots,d^*_k[n] ]= \E[\{\mathbf{b}_k \}] \nonumber \\
=&  [\E[\{b_k[1] \}],\E[\{b_k[2] \}],\ldots,\E[\{b_k[n] \}]] \nonumber \\
\overset{(c)} =& \underbrace{[\E[\mathcal{X}_k],\E[\mathcal{X}_k],\ldots,\E[\mathcal{X}_k]]}_{\text{length~$n$}} = \underbrace{[d^*_k,d^*_k,\ldots,d^*_k]}_{\text{length~$n$}}.
\end{align}
Thus, $\eta(\mathcal{C}-\mathbf{d})$ is the $n$-fold Cartesian product of one-dimensional constellation $\eta(\sqrt{\alpha}(\mathcal{X}_1-d^*_1)+\sqrt{1-\alpha}(\mathcal{X}_2-d^*_2))$ with rotation.
\end{IEEEproof}

With Lemma \ref{the:1}, we prove an upper bound on $d_{p,\min}(\eta(\mathcal{C}-\mathbf{d}))$ in the following.

\begin{lemma}\label{the:2a}
Consider a normalized and dithered superimposed constellation $\eta(\mathcal{C}-\mathbf{d})$ defined in \eqref{eq:sup1} for $\alpha \in [0,1]$ and $\Lambda$ is equivalent to $\mathbb{Z}^n$. The minimum product distance of $\eta(\mathcal{C}-\mathbf{d})$ is upper bounded by
\begin{align}
d_{p,\min}(\eta(\mathcal{C}-\mathbf{d})) \leq d_{p,\min}(\eta(\mathcal{X}-d^*)\mathbf{R}^*),
\end{align}
where $\eta(\mathcal{X}-d^*)$ is the one-dimensional constellation defined in \eqref{eq:1dsup}; and $\mathbf{R}^*$ is an $n \times n$ rotation matrix such that $d_{p,\min}(\eta(\mathcal{X}-d^*)\mathbf{R}^*) = d_{p,\min}(\Lambda)$ when $\alpha = 0$ or 1.

\end{lemma}
\begin{IEEEproof}
Given the definition of layer in Sec. \ref{def:layer} and based on Lemma \ref{the:1}, it is worth noting that all the layers have the same Euclidean distance profiles and thus their minimum Euclidean distances, denoted by $d_{E,\min}((\mathcal{X}-d^*)\mathbf{R})$, are the same regardless of any rotation $\mathbf{R}$. Thus,
\begin{align}\label{eq:dmin_1}
d_{E,\min}((\mathcal{X}-d^*)\mathbf{R}) = d_{E,\min}(\mathcal{X}-d^*).
\end{align}

When $\alpha = 0$ or 1, the superimposed constellation becomes a single user's constellation. In this case, the following relationship always holds
\begin{align}
d_{p,\min}((\mathcal{X}-d^*)\mathbf{R}) &\geq d_{p,\min}(\Lambda) = d_{p,\min}(\mathcal{C}-\mathbf{d}), \\
d_{E,\min}((\mathcal{X}-d^*)\mathbf{R}) &\overset{\eqref{eq:dmin_1}}= d_{E,\min}(\mathcal{X}-d^*) = d_{E,\min}(\Lambda), \alpha \in \{ 0,1\}.
\end{align}
Based on the above relationships and Lemma \ref{lem:dmindp} in Appendix~\ref{appendix:lemma}, there exists at least one layer such that the minimum product distance of this layer satisfies
\begin{align}\label{eq:pd_lambda_1}
d_{p,\min}((\mathcal{X}-d^*)\mathbf{R}^*) = d_{p,\min}(\Lambda) = d_{p,\min}(\mathcal{C}-\mathbf{d}), \alpha \in \{ 0,1\},
\end{align}
for some rotation matrix $\mathbf{R}^*$. By using \eqref{eq:dmin_1}-\eqref{eq:pd_lambda_1} and the relationship between minimum product distances in two different layers established in Lemma \ref{the:1a} in Appendix \ref{appendix:lemma}, we conclude that
\begin{align}
d_{p,\min}((\mathcal{X}-d^*)\mathbf{R}^*) \leq d_{p,\min}((\mathcal{X}-d^*)\mathbf{R}).
\end{align}

Now, we denoted by $d_{p,\min}(\mathcal{L})$ the minimum of the set of all product distances between all pairs of two distinct constellation points in any two \emph{different} layers. It is obvious that
\begin{align}
d_{p,\min}(\eta(\mathcal{C}-\mathbf{d})) &= \min\{d_{p,\min}(\eta\mathcal{L}) ,d_{p,\min}(\eta(\mathcal{X}-d^*)\mathbf{R}^*)\} \nonumber \\
&\leq d_{p,\min}(\eta(\mathcal{X}-d^*)\mathbf{R}^*),
\end{align}
where the normalize factor $\eta$ does not affect the equality and inequality here.
\end{IEEEproof}

With the upper bound in Lemma \ref{the:2a}, we now restrict the problem of bounding the minimum product distance of an $n$-dimensional constellation to analyzing the intra-layer minimum product distance $d_{p,\min}(\eta(\mathcal{X}-d^*)\mathbf{R}^*)$. This approach turns out to be sufficient for our purpose as it captures the trends of the change of the $d_{p,\min}(\eta(\mathcal{C}-\mathbf{d}))$ and fits perfectly with many local maximum values, as already shown in Example \ref{exa:dpmin}.

\begin{remark}\label{the:2}
When analyzing the minimum product distance of the superimposed constellation $\eta(\mathcal{C}-\mathbf{d})$, we only need to analyze the case for $\alpha \in [0,\frac{1}{2}]$. Specifically, the superimposed constellation  $\eta(\sqrt{\alpha}(\mathcal{C}_1-\mathbf{d}_1)+\sqrt{1-\alpha}(\mathcal{C}_2-\mathbf{d}_2))$ for $\alpha \in [\frac{1}{2},1]$ is equivalent to $\eta(\sqrt{\alpha'}(\mathcal{C}_2-\mathbf{d}_2)+\sqrt{1-\alpha'}(\mathcal{C}_1-\mathbf{d}_1))$ for $\alpha'  = 1 - \alpha \in [\frac{1}{2},0]$. Thus, the later case is analyzed when we let $m'_1 = m_2$ and $m'_2=m_1$ such that $\eta(\mathcal{C}-\mathbf{d}) = \eta(\sqrt{\alpha'}(\mathcal{C}'_1-\mathbf{d}'_1)+\sqrt{1-\alpha'}(\mathcal{C}'_2-\mathbf{d}'_2))$, where $\mathcal{C}'_k$ corresponds to the complete set of coset leaders of $\Lambda/2^{m'_k}\Lambda$ and $\mathbf{d}'_k = \E[\mathcal{C}'_k]$ for $k = 1,2$.
\end{remark}

Based on the definitions given in Sec. \ref{sec:def}, the intra-layer minimum product distance is
\begin{align}
d_{p,\min}(\eta(\mathcal{X}-d^*)\mathbf{R}^*) 
= \min \{d_{p,\min1},d_{p,\min2}, d_{p,\min}^{\text{Cl}}\}.
\end{align}
Since $d_{p,\min1}$ and $d_{p,\min2}$ can be easily computed as in \eqref{eq:dp1} and \eqref{eq:dp2}, respectively, what is left is to analyze $d_{p,\min}^{\text{Cl}}$. To perform the analysis, we first consider the case of $m_1 \in \mathbb{Z}^+$ and $m_2=1$ and then use the result to analyze the general case of $m_1,m_2 \in \mathbb{Z}^+$.

\subsubsection{Case I} ($m_2=1$)
For this case, there are two clusters, each of which contains $2^{m_1}$ number of constellation points. 
Before the constellation points from two clusters start to overlap, the inter-cluster minimum product distance is the product distance between two constellation points at the edge of each cluster. This scenario is illustrated in the example shown in Fig. \ref{fig:2}. The inter-cluster minimum product distance is given by
\begin{align}\label{eq:case1a}
\sqrt[n]{d_{p,\min}^{\text{Cl}}} =\sqrt[n]{d_{p,\min2}} -(2^{m_1}-1)\sqrt[n]{d_{p,\min1}},
\end{align}
where we have used the relationship of product distances in two line segments in $\mathbb{R}^n$ established in Lemma \ref{lem:1} in Appendix \ref{appendix:lemma}. We emphasize that Lemma \ref{lem:1} will be frequently used in the rest of the proof. Since $d_{p,\min 1} \leq d_{p,\min 2}$ for $\alpha \in [0,\frac{1}{2}]$ according to \eqref{eq:dp1} and \eqref{eq:dp2}, the intra-layer minimum product distance is thus determined by comparing $d_{p,\min}^{\text{Cl}}$ and $d_{p,\min 1}$. To have $d_{p,\min}^{\text{Cl}} \geq d_{p,\min 1}$, the necessary condition to satisfy this inequality is
\begin{align}
&\sqrt[n]{d_{p,\min2}} -(2^{m_1}-1)\sqrt[n]{d_{p,\min1}} \geq \sqrt[n]{d_{p,\min 1}} \nonumber \\
\Rightarrow & \; \sqrt[n]{(\sqrt{1-\alpha})^n}  \geq (2^{m_1}-1)\sqrt[n]{(\sqrt{\alpha})^n}  
\Rightarrow \; \alpha \leq \frac{1}{1+2^{2m_1}}
\end{align}
Thus, when $\alpha_1 \in [0,\frac{1}{1+2^{2m_1}}]$, we have
\begin{align}\label{eq:b4c}
d_{p,\min}(\eta(\mathcal{C}-\mathbf{d})) \leq d_{p,\min}(\eta(\mathcal{X}-d^*)\mathbf{R}^*) = d_{p,\min1}.
\end{align}
Then, for $\alpha \in (\frac{1}{1+2^{2m_1}},\frac{1}{2})$, the intra-layer minimum product distance becomes the inter-cluster minimum product distance such that
\begin{align}
d_{p,\min}(\eta(\mathcal{X}-d^*)\mathbf{R}^*) \overset{(a)}= d_{p,\min}^{\text{Cl}},
\end{align}
where $(a)$ follows that $d_{p,\min}^{\text{Cl}} < d_{p,\min1} < d_{p,\min2}$ for $\alpha \in (\frac{1}{1+2^{2m_1}},\frac{1}{2})$. Thus, we can now focus on analyzing $d_{p,\min}^{\text{Cl}}$ for this range.

To simplify the description for the subsequent analysis, we label two clusters as clusters 1 and 2, respectively, from the left to the right of a layer. Moreover, we refer to the Voronoi cell of an element with respect to the underlying rotated $\mathbb{Z}$ lattice in cluster 1 as a \emph{cell} of cluster 1. For each cluster, there are $(2^{m_1}-1)$ cells which are labelled cell $1$ to $(2^{m_1}-1)$, respectively, from the left to the right of a cluster. With $\alpha$ increasing, two clusters are moving toward each other. When the left constellation point on cell $(2^{m_1}-1)$ in cluster 1 overlaps with the right constellation point on cell 1 in cluster 2, $d_{p,\min}^{\text{Cl}} = 0$. From \eqref{eq:case1a}, this happens when
$\alpha = \frac{1}{(2^{m_1}-1)^2+1}$.
After the overlapping, the inter-cluster minimum product distance is bounded by
\begin{align}\label{eq:con1}
\sqrt[n]{d_{p,\min}^{\text{Cl}}}  \leq \frac{1}{2}\sqrt[n]{d_{p,\min1}},
\end{align}
where $\frac{1}{2}$ is due to the fact that the maximum of the inter-cluster product distance happens when a constellation point from cluster 2 is located in the center of a cell in cluster 1.

\begin{figure}[ht!]
	\centering
\includegraphics[width=3.43in,clip,keepaspectratio]{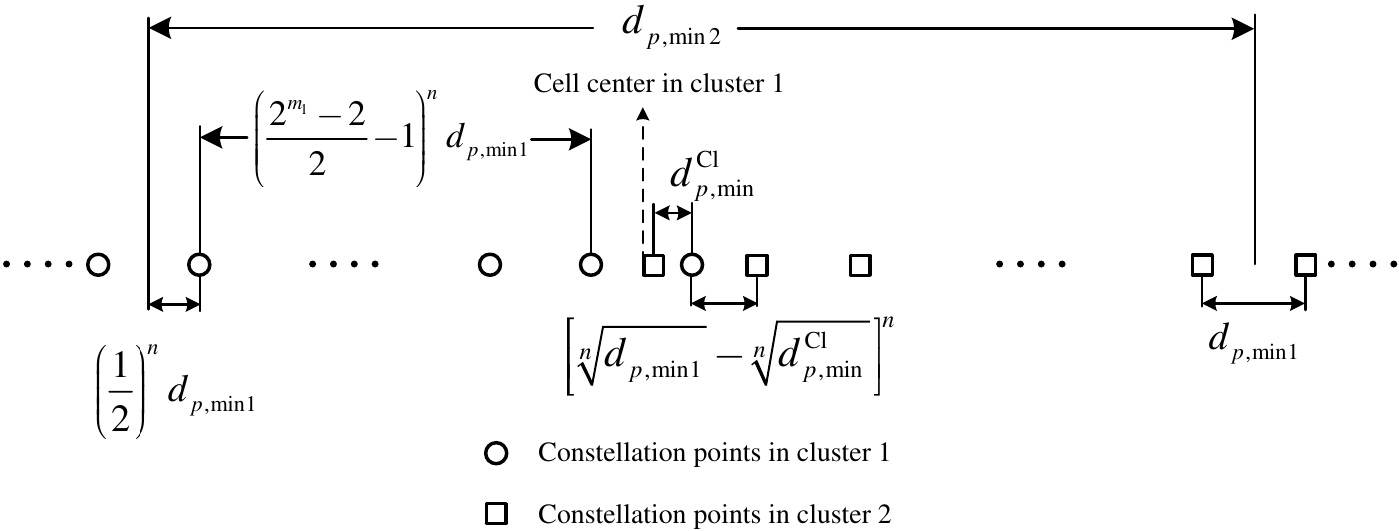}
\caption{An example of a layer in Case I.}
\label{fig:ex1}
\end{figure}

Consider the scenario where the leftmost constellation point of cell 1 of cluster 2 is in between the center and the right edge of cell $(2^{m_1}-1)$ in cluster 1. To have a clear view on this, we plot this scenario in Fig. \ref{fig:ex1}.

By counting the number of cells within clusters and inspecting the relationship between different product distances as shown in Fig. \ref{fig:ex1}, the inter-cluster minimum product distance is derived as
\begin{align}\label{eq:case1a2}
&2\left(\frac{1}{2}\sqrt[n]{d_{p,\min1}}\right)+2\left(\frac{2^{m_1}-2}{2}-1\right)\sqrt[n]{d_{p,\min1}} \nonumber \\
&+2\left(\sqrt[n]{d_{p,\min1}}-\sqrt[n]{d_{p,\min}^{\text{Cl}}}\right)+\sqrt[n]{d_{p,\min}^{\text{Cl}}} = \sqrt[n]{d_{p,\min2}} \nonumber \\
 \Rightarrow  \;& d_{p,\min}^{\text{Cl}} =\left( (2^{m_1}-1)\sqrt[n]{d_{p,\min1}} - \sqrt[n]{d_{p,\min2}}\right)^n.
\end{align}

Similarly, for the case where the left constellation point in cell 1 of cluster 2 is located in between the center of cell $(2^{m_1}-l)$ and cell $(2^{m_1}-l+1)$ in cluster 1, the inter-cluster minimum product distance is
\begin{align}\label{eq:con2a}
d_{p,\min}^{\text{Cl}} =\left| \sqrt[n]{d_{p,\min2}} -(2^{m_1}-l)\sqrt[n]{d_{p,\min1}}\right|^n, \; l = 2,\ldots,2^{m_1}-2.
\end{align}

By combining the right hand side of \eqref{eq:con1}, \eqref{eq:con2a}, the boundary of $\alpha$ corresponding to the $d_{p,\min}^{\text{Cl}}$ in \eqref{eq:con2a} can be computed as
\begin{align}
\alpha = \frac{1}{(2^{m_1}+\frac{1}{2}-l)^2+1}, \; l = 2,\ldots,2^{m_1}-2.
\end{align}
Summarizing the above results, we obtain the inter-cluster minimum product distance for case I in \eqref{eq:case1} for $\alpha \in (\frac{1}{1+2^{2m_1}},\frac{1}{2}]$.

\subsubsection{Case II} ($m_2\geq 1$)
First, it is obvious that the intra-layer minimum product distance is the same as in \eqref{eq:b4c} of Case I when $\alpha_1 \in [0,\frac{1}{1+2^{2m_1}}]$. However, \eqref{eq:case1} does not hold anymore when multiple clusters start to intercept. Since there are $2^{m_2}$ clusters, different constellation points from multiple clusters can be located in a cell of any cluster. Similar to Case I, we label all the clusters as $1,\ldots,2^{m_2}$ from the left cluster to the right cluster in a layer to simplify the description in the following analysis.

We denote by $\xi$ the number of clusters intercept with \emph{each other}, i.e., there are $\xi-1$ different constellation points from $ \xi-1$ different clusters, respectively, intercept with the cells of cluster 1. When $\xi = 2$, the scenario becomes identical to Case I and the same analysis on the minimum product distance applies. For $\xi \geq 3$, the inter-cluster minimum product distance takes into account that the cells of clusters $1,\ldots,\xi$ intercepting with each other. Thus, it can be bounded by
\begin{align}\label{eq:con3}
\sqrt[n]{d_{p,\min}^{\text{Cl}}(\xi)}   \leq \frac{1}{\xi} \sqrt[n]{d_{p,\min1}},
\end{align}
where $\frac{1}{\xi}$ comes from the same reason that we have $\frac{1}{2}$ in \eqref{eq:con1}. 
Now consider any two clusters $s,w \in \{2,\ldots,\xi \}$ and we assume $s>w$ without loss of generality. In the following, for the ease of presentation, we refer to $d_{p,\min}^{\text{Cl}}$ in case I as $d_{p,\min}^{\text{Cl}(m_2=1)}$. It can be easily seen that when $s-w = 1$, $d_{p,\min}(\text{Cl}_w,\text{Cl}_s)$ and $d_{p,\min}^{\text{Cl}(m_2=1)}$ coincide. For $s-w > 1$, to determine $d_{p,\min}(\text{Cl}_w,\text{Cl}_s)$, we first need to find a set of product distances between cluster 1 and $j$ for $j\in \{s,w\}$ as follows. 
Suppose that there are $F_j$ constellation points from cluster $j$ that have intercepted with the cells of cluster 1. By applying Lemma~\ref{lem:1} multiple times, we obtain the product distance between a point in cluster 1 and the $(f_j+1)$-th constellation point from cluster $j$ intercepting with the cells of cluster 1 (called $\text{Cl}_j(f_j)$) as
\begin{align}\label{eq:con4}
&\sqrt[n]{d_{p}(\text{Cl}_1,\text{Cl}_j(f_j))} = (j-1)\sqrt[n]{d_{p,\min}^{\text{Cl}(m_2=1)}} +f_j\sqrt[n]{d_{p,\min 1}}, \nonumber \\
&f_j \in \{ 0,\ldots,F_j-1 \},j \in \{s,w\},
\end{align}
The minimum product distance $d_{p,\min}(\text{Cl}_w,\text{Cl}_s)$ is then computed as
\begin{align}\label{eq:dpij}
&d_{p,\min}(\text{Cl}_w,\text{Cl}_s) \nonumber \\
\overset{(a)}{=}& \min\limits_{\substack{f_w \in \{ 0,\ldots,F_w-1 \}, \\ f_s \in \{ 0,\ldots,F_s-1 \}}}\left\{\left|\sqrt[n]{d_{p}(\text{Cl}_1,\text{Cl}_w(f_w))} -\sqrt[n]{d_{p}(\text{Cl}_1,\text{Cl}_s(f_s))} \right|^n \right\}\nonumber \\
\overset{\eqref{eq:con4}} =& \min\limits_{\substack{f_w \in \{ 0,\ldots,F_w-1 \}, \\ f_s \in \{ 0,\ldots,F_s-1 \}}} \bigg\{\bigg|(f_w-f_s)\sqrt[n]{d_{p,\min 1}} \nonumber \\
&+(w-1)\sqrt[n]{d_{p,\min}^{\text{Cl}(m_2=1)}} 
- (s-1)\sqrt[n]{d_{p,\min}^{\text{Cl}(m_2=1)}}\bigg|^n \bigg\} \nonumber \\
 \overset{(b)}=& \min\limits_{\gamma_{ws}\in \{0,\ldots,\lfloor\frac{s-w}{2} \rfloor\} }\left\{ \left|\gamma_{ws}\sqrt[n]{d_{p,\min1}}  
-(s-w)\sqrt[n]{d_{p,\min}^{\text{Cl}(m_2=1)}}\right|^n  \right\},
\end{align}
where $(a)$ follows from Lemma~\ref{lem:1} and $(b)$ follows from the fact that $\gamma_{ws} $ is the spacing between cluster $s$ and $w$ in terms of $\sqrt[n]{d_{p,\min1}}$  and $\lfloor\frac{s-w}{2} \rfloor$ is the maximum spacing because
\begin{align}
(s-w)\sqrt[n]{d_{p,\min}^{\text{Cl}(m_2=1)}} \overset{\eqref{eq:con1}}\leq  \frac{s-w}{2}\sqrt[n]{d_{p,\min 1}}.
\end{align}

For $\xi \in \{2, \ldots,2^{m_2}\}$, the inter-cluster minimum product distance for the scenario where the cells of clusters $1,\ldots,\xi$ intercept with each other, is obtained by finding the minimum of the product distances based on all combinations of clusters $s$ and $w$
\begin{align}
&d_{p,\min}^{\text{Cl}}(\xi)= \min_{\substack{s,w \in \{1,\ldots,\xi \},\\ s > w}} \left\{ d_{p,\min}(\text{Cl}_w,\text{Cl}_s) \right\}\nonumber \\
 = &\min_{\substack{w,s \in \{1,\ldots,\xi \},s > w \\ \gamma_{ws}\in \{0,\ldots,\lfloor\frac{s-w}{2} \rfloor\}}} \left\{\left|\gamma_{ws}\sqrt[n]{d_{p,\min 1}}
-(s-w)\sqrt[n]{d_{p,\min}^{\text{Cl}(m_2=1)}}\right|^n  \right\} \nonumber \\
 \overset{(a)}=& \min\limits_{\substack{\gamma \in \{ 0,\ldots,\lfloor\frac{\xi-1}{2} \rfloor \}, \\ \beta \in \{1,\ldots,\xi-1\}}} \left\{\left|\gamma\sqrt[n]{d_{p,\min 1}}
-\beta\sqrt[n]{d_{p,\min}^{\text{Cl}(m_2=1)}}\right|^n  \right\},
\end{align}
where $(a)$ follows from that $1 \leq s-w \leq \xi -1$.

The only thing left is to find the  boundary of $\alpha$, called $\alpha(\xi)$, such that when $\alpha \geq \alpha(\xi)$, the bound in \eqref{eq:con3} is valid. This happens when the minimum product distance between cluster 1 and cluster $\xi$ satisfies the condition of $\sqrt[n]{d_{p,\min}(\text{Cl}_1,\text{Cl}_{\xi})}\leq \frac{1}{2}\sqrt[n]{d_{p,\min1}}$. Otherwise, the above scenario is reduced to the scenario of the cells of clusters $1,\ldots,(\xi-1)$ intercepting with each other because $\sqrt[n]{d_{p,\min}(\text{Cl}_1,\text{Cl}_{\xi})}> \frac{1}{2}\sqrt[n]{d_{p,\min1}} \geq \sqrt[n]{d_{p,\min}^{\text{Cl}(m_2=1)}}$ leads to
\begin{align}
d_{p,\min}^{\text{Cl}}(\xi) &= \min \{d_{p,\min}(\text{Cl}_1,\text{Cl}_{\xi}),\ldots, d_{p,\min}(\text{Cl}_{\xi-1},\text{Cl}_{\xi})\} \nonumber \\
&\overset{(a)}= \min \{d_{p,\min}(\text{Cl}_{2},\text{Cl}_{\xi}),\ldots, d_{p,\min}(\text{Cl}_{\xi-1},\text{Cl}_{\xi})\} \nonumber \\
& \overset{(b)}= \min \{d_{p,\min}(\text{Cl}_{1},\text{Cl}_{\xi-1}),\ldots, d_{p,\min}(\text{Cl}_{\xi-2},\text{Cl}_{\xi-1})\} \nonumber \\
&= d_{p,\min}^{\text{Cl}}(\xi-1),
\end{align}
where $(a)$ follows from that $d_{p,\min}(\text{Cl}_{\xi-1},\text{Cl}_{\xi}) = d_{p,\min}^{\text{Cl}(m_2=1)}<d_{p,\min}(\text{Cl}_{1},\text{Cl}_{\xi})$ and $(b)$ follows from \eqref{eq:dpij} that $d_{p,\min}(\text{Cl}_{w_2},\text{Cl}_{s_2}) = d_{p,\min}(\text{Cl}_{w_2},\text{Cl}_{s_2})$ if $s_1 - w_1 = s_2-w_2$ for any $s_1,w_1,s_2,w_2 \in \{1,\ldots,\xi \}$. Thus, the corresponding $\alpha(\xi)$ is derived by using the above condition as
\begin{align}
&(\xi-1)\sqrt[n]{d_{p,\min 2}} = \left(2^{m_1}-\frac{1}{2}\right)\sqrt[n]{d_{p,\min1}} \nonumber \\
\Rightarrow \; &\alpha(\xi) = \frac{(\xi-1)^2}{\left(2^{m_1}-\frac{1}{2} \right)^2+(\xi-1)^2}.
\end{align}
Note that we only needs to look at $\alpha(\xi) \leq \frac{1}{2}$ according to Remark \ref{the:2}. This completes the proof.

\section{Extension to MIMO-NOMA}\label{MIMO}
In this section, we extend the main idea and analysis to MIMO-NOMA over block fading channels for constructing good MIMO-NOMA schemes without SIC. We restrict our attention to a very popular class of codes for MIMO channel named OSTBC. Some advantages of using OSTBC include achieving full transmit diversity and efficiently detection by turning the MIMO channel into a set of non-interfering parallel subchannels. We note that a scheme of NOMA with two transmit antennas and one receive antenna for each user combined with Alamouti code \cite{730453} has been reported in \cite{8392409} where the closed form expressions for outage probabilities under Nakagami-m fading channels are derived. However, the analysis is based on Gaussian inputs. In this section, we adapt the techniques used in the previous section to analyze the error performance of MIMO-NOMA scheme with general OSTBC \cite{Vucetic:2003:SC:861866}.

\subsection{MIMO-NOMA system model}
Consider a two-user MIMO-NOMA where the base station and each user have $M_t$ and $M_r$ sufficient-spacing antennas, respectively. We again assume that the transmitter has statistical CSI while the receiver has full CSI for its own channel. The base station encodes all users' messages $\mathbf{u}_1,\mathbf{u}_2$ into a superimposed codeword $\mathbf{X} = [\mathbf{x}_1,\ldots,\mathbf{x}_T] \in \mathbb{C}^{M_t \times T}$ from the codebook $\mathcal{G}$ and broadcasts it to each user, where $T$ means the codeword spreads $T$ time slots and $\sum_{i=1}^T\E[\|\mathbf{x}_i\|^2]\leq T$ for $i = 1,\ldots,T$. We denote by $\mathbf{H}_k \in \mathbb{C}^{M_r \times M_t}$ the channel matrix for user $k \in \{1,2\}$ with i.i.d. entries. Here, we assume that $\mathbf{H}_k$ is constant during one codeword block while a transmit packet contains multiple blocks. The received signal at user $k$ for $T$ time slots is denoted by $\mathbf{Y}_k \in \mathbb{C}^{M_r \times T}$ and is given by
\begin{align}\label{STBC_Y}
\mathbf{Y}_k = \sqrt{P}\mathbf{H}_k\mathbf{X}+\mathbf{Z}_k,
\end{align}
where $P$ is the total power constraint at the base station and $\mathbf{Z}_k \in \mathbb{C}^{M_r \times T}$ is a circular-symmetric AWGN experienced at user $k$ with i.i.d. entries $\sim\mc{CN}(0,1)$. The reliability is again measured by PEP. Consider the channels $\mathbf{H}_k$ with i.i.d. entries $h_{j,i}^{(k)}\sim\mc{CN}(0,\sigma_k^2)$. Following \cite[Ch. 2.5.1]{Vucetic:2003:SC:861866}, for any two codewords $\mathbf{X}_s,\mathbf{X}_w \in \mathcal{G}$ and $\mathbf{X}_s \neq \mathbf{X}_w$, user $k$'s error probability is upper bounded by its average PEP given by
\begin{align}\label{PEP_MIMO_NOMA}
P_e^{(k)} &\leq \frac{1}{|\mathcal{G}|}\sum_{s \neq w}\text{Pr}(\mathbf{X}_s \rightarrow \mathbf{X}_w|\mathbf{H}_k) \nonumber \\
&\leq \frac{1}{|\mathcal{G}|}\sum_{s \neq w}\det\left(\mathbf{I}_{M_t}+\overline{\SNR}_k\frac{(\mathbf{X}_s-\mathbf{X}_w)(\mathbf{X}_s-\mathbf{X}_w)^{\dag}}{4}\right)^{-M_r} \nonumber \\
&\leq \frac{1}{|\mathcal{G}|}\sum_{s \neq w} \prod_{j=1}^{r}\left(1+ \overline{\SNR}_k\frac{\phi_j}{4}\right)^{-M_r} \nonumber \\ 
&< \left(\min\limits_{s \neq w}\left\{\prod\nolimits_{j=1}^{\min\limits_{s\neq w}\{r\}} \phi_j\right\}\right)^{-Mr}\left(\frac{4}{\overline{\SNR}_k}\right)^{\min\limits_{s\neq w} \{r\}M_r}  ,
\end{align}
where $(.)^{\dag}$ denotes the conjugate transpose, $\overline{\SNR}_k \triangleq \E[\text{tr}(\mathbf{H}_k\mathbf{H}_k^{\dag})]P$ is user $k$'s average SNR; $\{ \phi_j:j = 1,\ldots,r\}$ are the non-zero eigenvalues of $\Delta\Delta^{\dag}$ with $\Delta \triangleq \mathbf{X}_s-\mathbf{X}_w$ being the codeword difference matrix with $r = \text{rank}(\Delta)$. The diversity order of $\mathbf{X}$ is $\min\limits_{s\neq w} \{r\}M_r$. For $T \geq M_t$, the code has full rank such that $\min\limits_{s\neq w} \{r\} = M_t$ and $\prod_{j=1}^{M_t}\phi_j = \det(\Delta\Delta^{\dag})$. In this case, the code achieves full diversity, i.e., the diversity order is $M_tM_r$. To further minimize the PEP, it is important to maximize the \emph{minimum determinant} $\min\limits_{s\neq w}\{\det(\Delta\Delta^{\dag})\}$. It is noteworthy that the design criterion is generalizable to other fading channels, e.g., Rician fading \cite{Vucetic:2003:SC:861866,1256737}. Without loss of generality, we assume that $\overline{\SNR_1}\geq \overline{\SNR_2}$ and user 1 is considered as the strong user. Note that this user ordering is also adopted in \cite{8392409}.

\subsection{Proposed scheme and main result}\label{sec:MIMO_scheme}

We first briefly describe the scheme of space-time coded MIMO-NOMA in the following.

\subsubsection{Transmitter side}\label{MIMO_NOMA_TX}
A superimposed signal sequence $[x[1],\ldots,x[M_t]]$ is encoded into a OSTBC codeword $\mathbf{X}\in \mathbb{C}^{M_t \times T}$, where $x[l] \in \eta_T(\mathcal{C}-\mathbf{d})$ can be expressed by \eqref{eq:sup1} for $n=2 ,l = 1,\ldots,M_t$ and $\eta_T = \tau \eta$ is applied to $\mathbf{X}$ to ensure $\sum_{i=1}^T\E[\|\mathbf{x}_i\|^2]\leq T$. Here $\tau$ is an additional normalization for the space-time code on top of the normalization of the superimposed constellation $\eta$ and it depends on the specific space-time code (for example, Alamouti code has $\tau = 1$). 

\subsubsection{Receiver side}
Upon receiving $\mathbf{Y}_k$ given in \eqref{STBC_Y}, the maximum-ratio combining and the space-time decoding are employed for decoding. By using the orthogonality of pairwise rows of the transmission matrix \cite[Ch 3.6]{Vucetic:2003:SC:861866}, the decoder attempts to minimize the following metric
\begin{align}
\arg\min_{\tilde{x}[l] \in \mathcal{G}}& \left\{|\tilde{x}[l]-x[l]|^2+\left(\sum\nolimits_{i = 1}^{M_t}\sum\nolimits_{j = 1}^{M_r}|h_{j,i}^{(k)}|^2-1\right)|x[l]|^2 \right\}, \nonumber\\
& l \in \{1,\ldots,M_t\},
\end{align}
where $\tilde{x}[l]$ is the estimated superimposed symbol of $x[l]$. If single-user decoding is adopted, each user directly decodes their own messages 
from $\tilde{x}[l]$ for $l = 1,\ldots,M_t$ in a symbol-wise manner. For user 1 with SIC user 2's message will be decoded from $\tilde{x}[l]$ first and the corresponding codeword will be re-encoded and subtracted from the received signal.

For lattice-partition based MIMO-NOMA scheme, the superimposed signal $x[l] \in  \eta_T'(\mathcal{C}-\mathbf{d}')$ with $\eta'_T = \tau \eta'$ can be expressed by \eqref{eqn:x_def}, which corresponds to the lattice partition chain described in Section \ref{Sec:LP}. As a result, many nice properties of the underlying lattice carry over to the individual and the superposition coded space-time codewords.

The analytical expression of the minimum Euclidean distance is similar to that of the minimum product distance given in \eqref{eq:case2}-\eqref{eq:case1}. We summarize the main result here and the proof is presented in Section~\ref{sec:MIMO_proof}.
\begin{proposition}\label{coro:dmin}
For arbitrary power allocation, the minimum determinant is $\min\{\det(\Delta\Delta^{\dag})\} = d_{E,\min}(\eta_T(\mathcal{C}-\mathbf{d}))^{2M_t}$, where $ d_{E,\min}(\eta_T(\mathcal{C}-\mathbf{d}))$ is obtained by replacing $d_{E,\min1}$ from \eqref{eq:dmin1} to $d_{p,\min1}$ and $d_{E,\min2}$ from \eqref{eq:dmin2} to $d_{p,\min2}$ and substituting them into \eqref{eq:case2}-\eqref{eq:case1} and setting $n=1$. 
For the lattice-partition based scheme, the minimum determinant is $\min\{\det(\Delta\Delta^{\dag})\} = d_{E,\min}(\eta_T'(\mathcal{C}'-\mathbf{d}'))^{2M_t}$, where $d_{E,\min}(\eta_T'(\mathcal{C}'-\mathbf{d}')) = \tau\eta'd_{E,\min}(\Lambda)$ and $\eta'$ is given in \eqref{normalize_LP}.
\end{proposition}

We would like to emphasize that although we present the results for the complex OSTBC over complex MIMO setting, the above results are valid for real OSTBC.
\begin{figure}[ht!]
	\centering
\includegraphics[width=3.4in,clip,keepaspectratio]{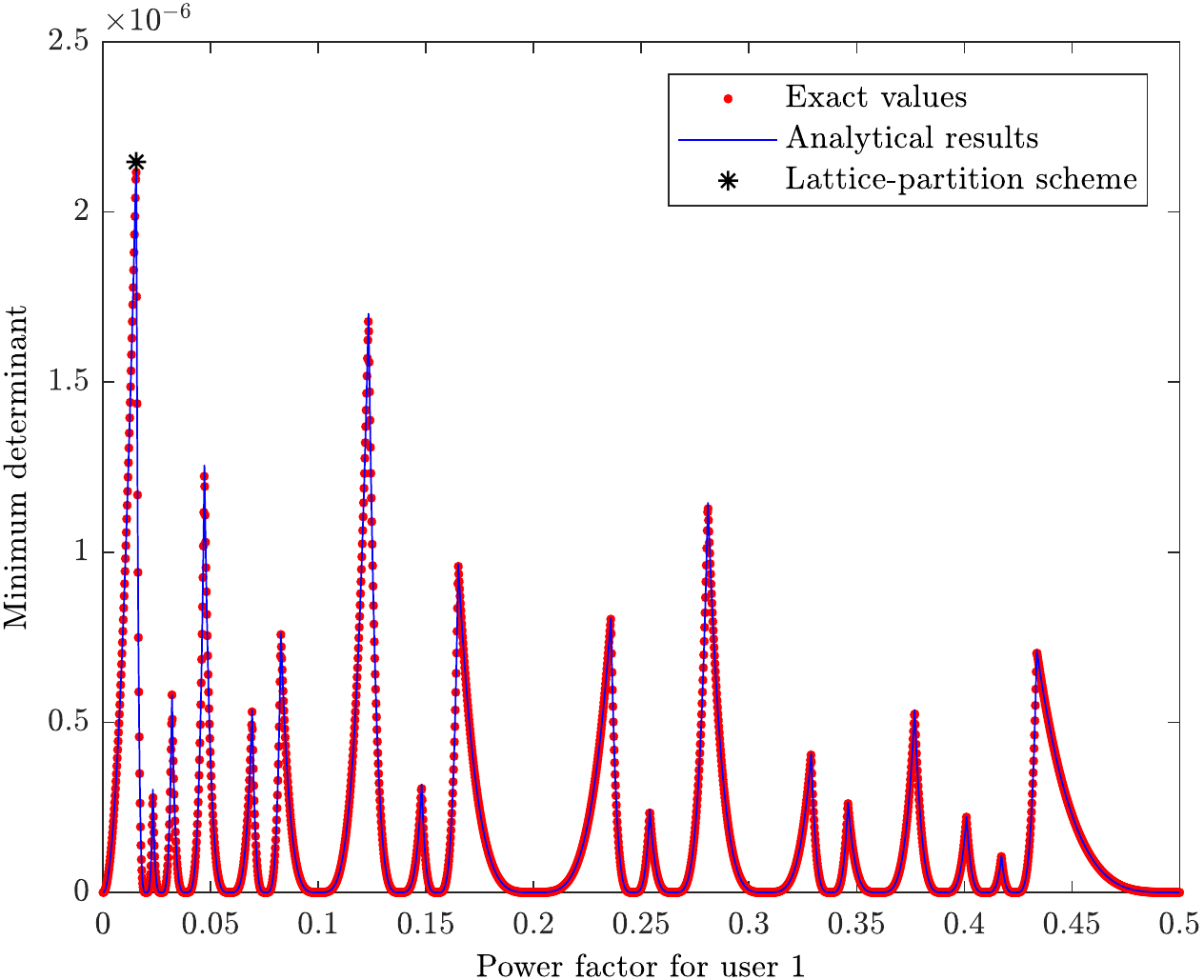}
\caption{Minimum determinant of Alamouti coded two-dimensional superimposed constellation from $(m_1,m_2) = (3,3)$}
\label{fig:14}
\end{figure}

We now give an example in the following to show the analytical results. We use the same setting as in Example \ref{exa:dpmin} and employ Alamouti code. We then obtain the exact values for $\min\{\det(\Delta\Delta^{\dag})\}$ by exhaustive search and compute the analytical results in Proposition \ref{coro:dmin}. The results are shown in Fig. \ref{fig:14}, from which one can observe that the analytical results perfectly match with the exact values of $\min\{\det(\Delta\Delta^{\dag})\}$. Moreover, the scheme based on lattice partition is again optimal in terms of minimum determinant.

\subsection{Proof of Proposition \ref{coro:dmin}}\label{sec:MIMO_proof}
Consider the transmission scheme described in Sec. \ref{sec:MIMO_scheme}. According to the design criterion for OSTBC \cite[Ch. 3.5]{Vucetic:2003:SC:861866}, the codeword matrix satisfies
$\mathbf{X}\mathbf{X}^{\dag} = (|x[1]|^2+\ldots+|x[M_t]|)\mathbf{I}_{M_t}$.
Hence, the minimum determinant of codeword difference matrix $\Delta \triangleq \mathbf{X}_s-\mathbf{X}_w$ is 
\begin{align}
&\min\limits_{s\neq w}\{\det(\Delta\Delta^{\dag})\} \nonumber \\
=& \min\limits_{s\neq w}\{\det((|x_s[1]-x_w[1]|^2+\ldots+|x_s[M_t]-x_w[M_t]|)\mathbf{I}_{M_t})\} \nonumber \\
 =& \min\limits_{s\neq w}\{(|x_s[1]-x_w[1]|^2+\ldots+|x_s[M_t]-x_w[M_t]|)^{M_t}    \} \nonumber \\
 \geq& \min\limits_{l \in \{1,\ldots,M_t \}} \{ \min_{s \neq w}\{|x_s[l]-x_w[l]|^{2M_t}\} \} \nonumber \\
 =& d_{E,\min}(\eta_T(\mathcal{C}-\mathbf{d}))^{2M_t}.
\end{align}

From this point onward, the problem of analyzing minimum determinant is reduced to that of analyzing the minimum Euclidean distance of the composite constellation $\eta_T(\mathcal{C}-\mathbf{d})$. 
In what follows, we prove the following lemma about the exact minimum Euclidean distance, which in turn, gives us the exact minimum determinant.

\begin{lemma}\label{the:dmin}
Consider the constellation $\eta(\mathcal{C}-\mathbf{d})$ defined in \eqref{eq:sup1} for $\alpha \in [0,1]$ and $\Lambda$ is equivalent to $\mathbb{Z}^n$. Then the minimum Euclidean distance of $\eta(\mathcal{C}-\mathbf{d})$ is
\begin{align}\label{eq:dmin_the}
d_{E,\min}(\eta(\mathcal{C}-\mathbf{d})) = d_{E,\min}(\eta(\mathcal{X}-d^*)).
\end{align}
\end{lemma}
\begin{IEEEproof}
From Lemmas \ref{the:1}-\ref{the:2a}, we know that each layer has the same Euclidean distance profile and thus same minimum Euclidean distance regardless of rotation. Similar to Lemma \ref{the:2a}, we have
\begin{align}
d_{E,\min}(\eta(\mathcal{C}-\mathbf{d})) = \min\{d_{E,\min}(\eta\mathcal{L}) ,d_{E,\min}(\eta(\mathcal{X}-d^*))\},
\end{align}
where $d_{E,\min}(\eta\mathcal{L})$ denotes the minimum of the set of Euclidean distances between all pairs of two distinct constellation points in any two different layers.

For any pair of non-intercepting layers from $\eta(\mathcal{C}-\mathbf{d})$, the minimum Euclidean distance between them is the length of the line segment that is orthogonal to these layers. The end points of this line segment are in fact the constellation points of a layer that is orthogonal to these layers. For any pair of intercepting layers within the composite constellation, the crossing point and two constellation points (each one is from different layer) form a right triangle. The Euclidean distance between these two constellation points is strictly larger than the Euclidean distance between the crossing point and either of those two constellation points, respectively. Thus, we conclude that
$d_{E,\min}(\eta\mathcal{L}) = d_{E,\min}(\eta(\mathcal{X}-d^*))$.
This completes the proof of \eqref{eq:dmin_the}.
\end{IEEEproof}

With Lemma \ref{the:dmin}, we obtain 
$\min\limits_{s\neq w}\{\det(\Delta\Delta^{\dag})\} = d_{E,\min}(\eta_T(\mathcal{X}-d^*))^{2M_t}$
by replacing the scalar with $\eta_T$.
To analyze this minimum Euclidean distance, we first denote by $d_{E,\min1}$ and $d_{E,\min2}$ the minimum Euclidean distance of constellation $\eta_T\sqrt{\alpha}(\mathcal{X}_1-d^*_1)$ and $\eta_T\sqrt{1-\alpha}(\mathcal{X}_2-d^*_2)$, respectively, where
\begin{align}
&d_{E,\min1}  \triangleq \eta_T\sqrt{\alpha} d_{E,\min}(\Lambda), \label{eq:dmin1} \\
&d_{E,\min2}  \triangleq \eta_T\sqrt{1-\alpha} d_{E,\min}(\Lambda). \label{eq:dmin2}
\end{align}
Here, $d_{E,\min}(\Lambda) = 1$ when the base lattice $\Lambda$ is equivalent to $\mathbb{Z}^n$ and $\eta_T = \tau\eta$, where $\eta$ is given in \eqref{normalize2}. Then, following the steps of our analysis in Sec. \ref{sec:dminp_ana} completes the proof. 

\section{Simulation Results}\label{SIM}
In this section, we provide the simulation results of our proposed scheme introduced in Sections~\ref{sec:proposed} and \ref{MIMO} and compare them with the current state-of-the-art.

\subsection{Single antenna case}\label{sim_SISO}
\begin{figure}[ht!]
	\centering
\includegraphics[width=3.21in,clip,keepaspectratio]{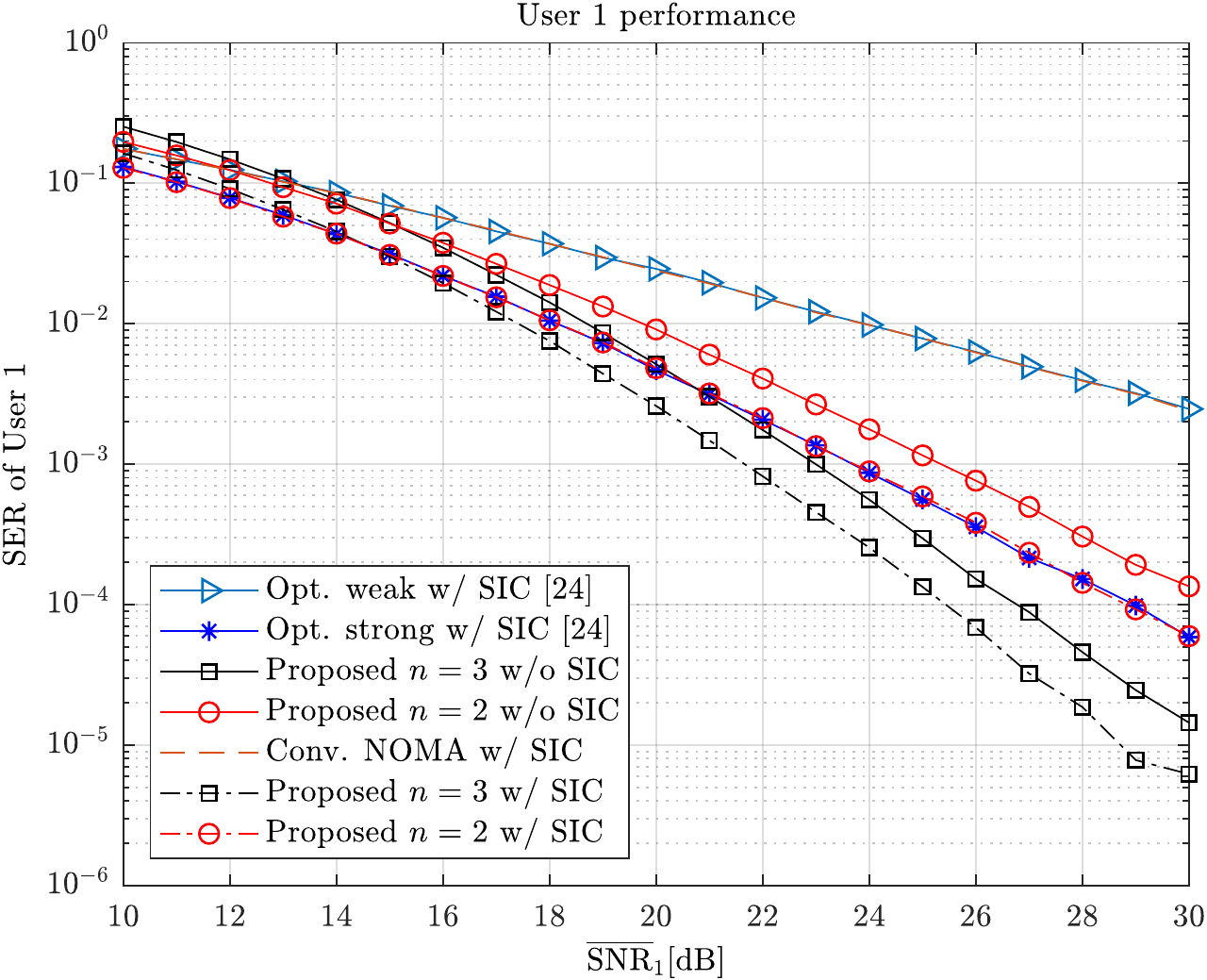}
\caption{Simulation results for user 1's SER.}
\label{fig:8a}
\end{figure}

\begin{figure}[ht!]
	\centering
\includegraphics[width=3.21in,clip,keepaspectratio]{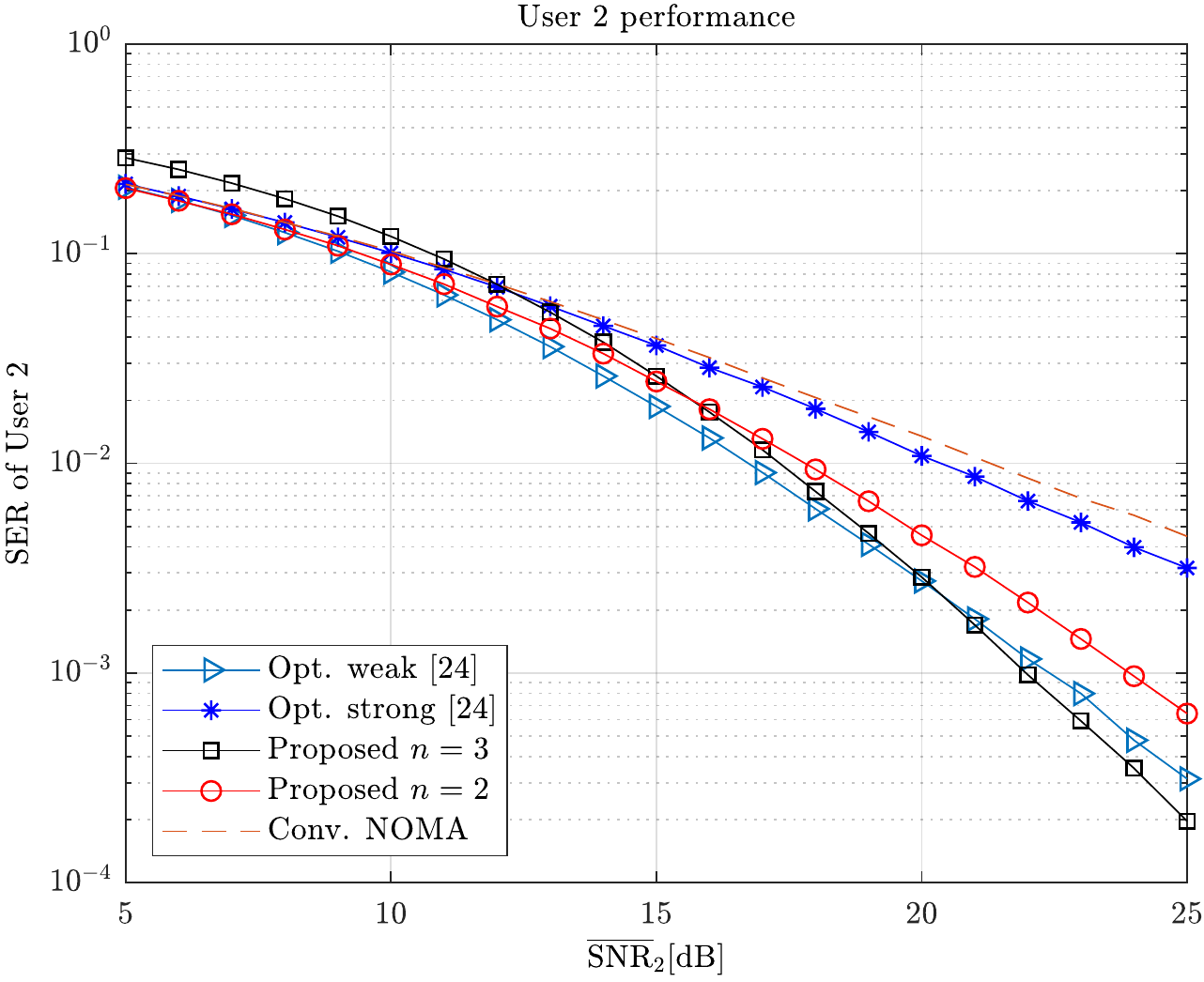}
\caption{Simulation results for user 2's SER.}
\label{fig:8b}
\end{figure}

In this subsection, we first provide simulation results of the lattice partitioned scheme for the single antenna case. The dimension of the underlying ideal lattice is set to $n= 2,3$. For illustrative purpose, we consider $(m_1,m_2) = (1,1)$ in order to make fair comparison with the scheme in \cite{7880967}. 
We use the conventional NOMA (labelled Conv. NOMA) scheme which adopts square 4-QAM (not rotated) as a benchmark. The performance of strong user (user 1) and that of weak user (user 2) are measured in terms of SER versus their average SNRs and plotted in Fig. \ref{fig:8a} and Fig. \ref{fig:8b}, respectively. In addition, the SER of the schemes in \cite{7880967} are plotted in both figures. Note that \cite{7880967} has \emph{two schemes} corresponding to optimization for strong user and optimization for weak user, respectively. We also emphasize here that the power allocations for the conventional NOMA scheme, our schemes and the schemes in \cite{7880967} are the same, i.e., $\alpha = 0.2$. In all the curves in these figures, when SIC is adopted at user 1, we assume that user 2's signals are perfectly decoded and subtracted.

It can be observed that for the proposed schemes with $n=2$ and 3, respectively, the full diversity orders of $2$ and 3, respectively, can be achieved for both users even without SIC. In particular, each user in our scheme for $n=2$ achieves comparable performance compared to the user whose constellation is optimized in \cite{7880967}.
Conversely, in \cite{7880967}, the performance at the user which is not optimized reveals no diversity gain as the conventional NOMA scheme. 
Furthermore, the maximum diversity order in scheme \cite{7880967} is only 2 while our scheme can provide higher diversity order and coding gain to both users by choosing higher-dimensional ideal lattices as the base lattices. Last but not least, our proposed scheme based on lattice partition provides a systematic way to design downlink NOMA scheme that offers full diversity gain and high coding gain, while the scheme in \cite{7880967} is based on exhaustive search.

\subsection{Multiple antennas case}
\begin{figure}[ht!]
	\centering
\includegraphics[width=3.21in,clip,keepaspectratio]{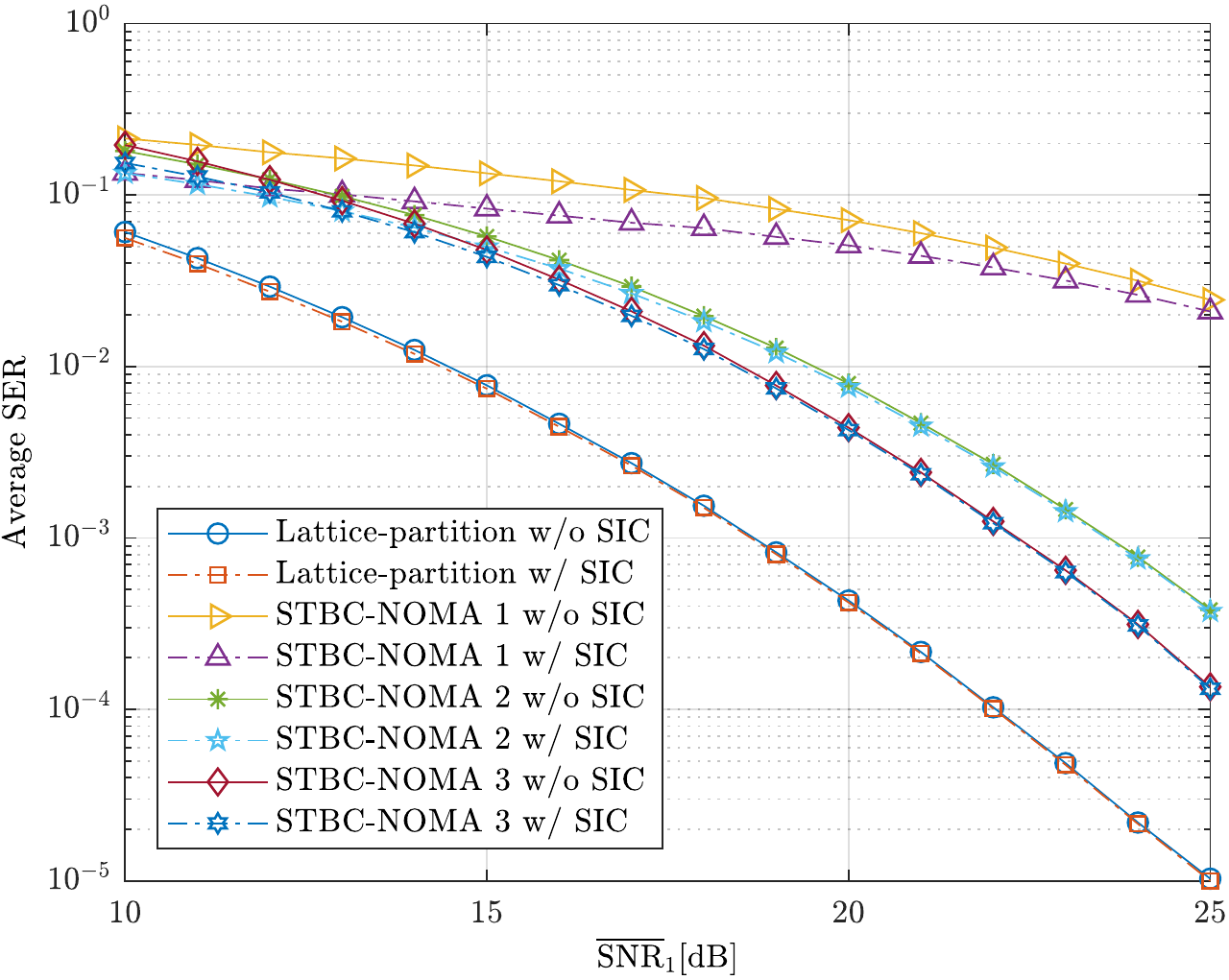}
\caption{Simulation results for average SER among two users.}
\label{fig:avg_SER}
\end{figure}

\begin{figure}[ht!]
	\centering
\includegraphics[width=3.21in,clip,keepaspectratio]{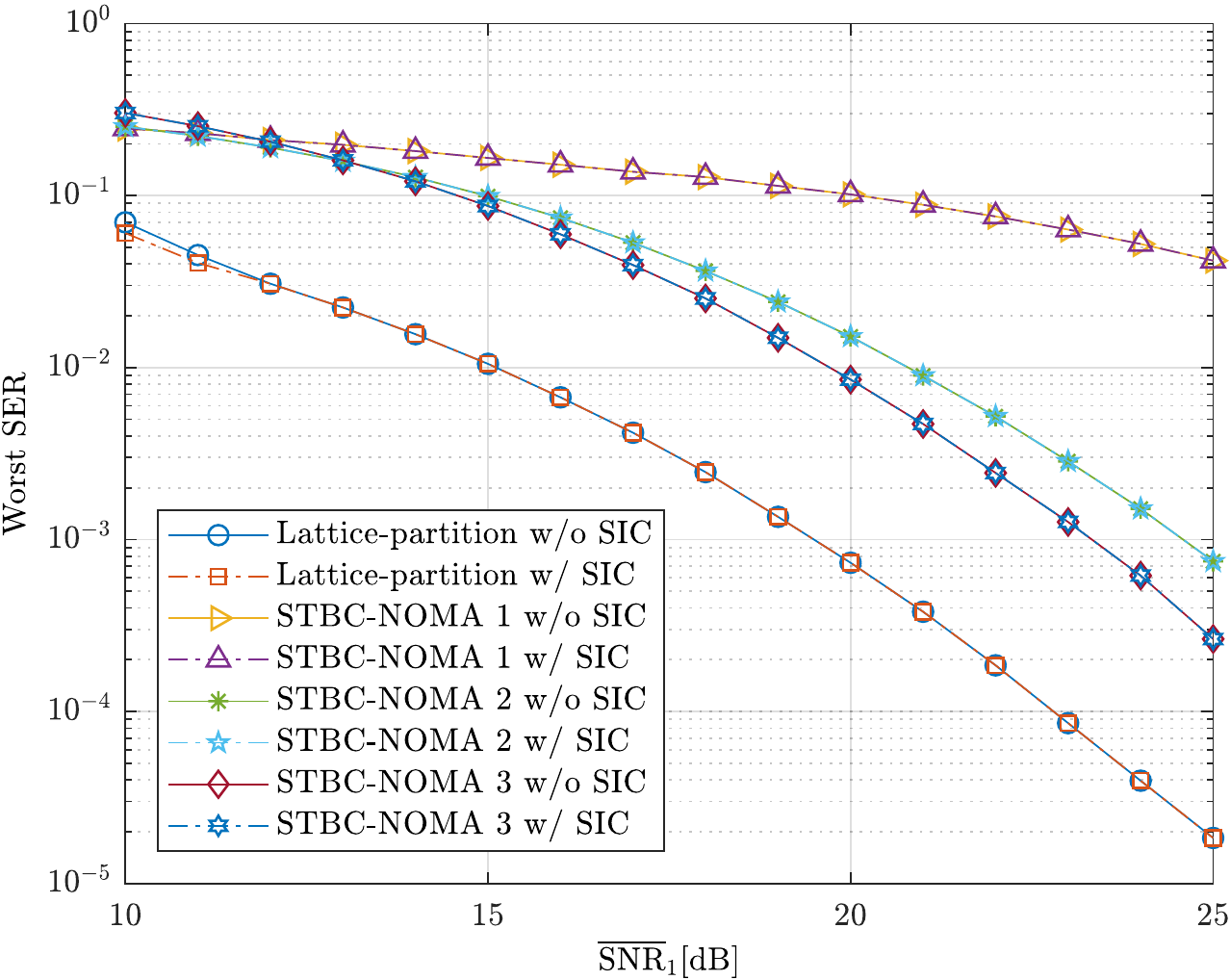}
\caption{Simulation results for worst SER among two users.}
\label{fig:max_SER}

\end{figure}

In this subsection, we provide the simulation results for the proposed MIMO-NOMA scheme where the base station and each user have two antennas and the underlying OSTBC is Alamouti code. We consider the case for $(m_1,m_2) = (2,1)$ and the channel is Rayleigh fading. The difference between $\overline{\text{SNR}}_1$ and $\overline{\text{SNR}}_2$ is 5 dB. Since we are unable to find a benchmark downlink MIMO-NOMA scheme with discrete inputs and with similar channel assumptions as ours, we thus compare the error performances of our lattice-partition scheme and a number of space-time block coded NOMA schemes with some power allocations.
Specifically, we choose $\alpha = 0.11,0.14$ and 0.31 for three schemes (labelled as STBC-NOMA 1-3, respectively) and the corresponding minimum determinants are $0.136 \cdot 10^{-4}$, $0.169 \cdot 10^{-2}$ and $0.449 \cdot 10^{-2}$, respectively. The lattice partition scheme has a minimum determinant of $0.91 \cdot 10^{-2}$. Here, the error performances are measured by average SER and worst case SER among two users versus user 1's average SNR. These results are plotted in Fig. \ref{fig:avg_SER} and Fig. \ref{fig:max_SER}, respectively. 

It can be seen that the scheme with larger minimum determinant has better error performance than that of the scheme with smaller minimum determinant. Another interesting observation is that the schemes with SIC only provide negligible gain for both average and the worst SER performance among two users. This is due to the fact that the
average/worst SER performance is largely dominated by the performance of the user with much higher SER.

\section{Concluding Remarks}\label{sec:conclude}
In this work, we have proposed a class of downlink NOMA scheme without SIC for block fading channels. In particular, we have used algebraic lattices to design modulations such that full diversity gain and large coding gain can be attained for all users at the same time. Moreover, the minimum product distance for the superimposed constellation for arbitrary power allocation has been thoroughly investigated. Within the proposed class, a family of schemes based on lattice partitions has then been identified. It has been shown via numerical result that schemes from this special family achieve the largest minimum product distances among the proposed class. An extension of the proposed scheme to the MIMO-NOMA system with OSTBC has then been introduced. The exact minimum determinant of the proposed scheme has been derived. Simulation results have been provided, which confirms our analytical results and also demonstrates that our schemes significantly outperform the current state-of-the-art.

\appendices
\section{Useful lemmas}\label{appendix:lemma}
\begin{lemma}\label{lem:zn_P}
Let $\msf{V}$ be a discrete random variable uniformly distributed over the complete set of coset leaders of $\Lambda/2^m\Lambda$ with any $m \in \mathbb{Z}^+$. Let $\mathbf{d} = \E[\msf{V}]$ be a dither vector such that $\msf{X} = \msf{V} - \mathbf{d}$ has zero mean. When $\Lambda = \mathbb{Z}^n$, the average power of $\msf{X}$ is given by
\begin{align}\label{eq:zn_P}
\E [\| \msf{X}\|^2] = \frac{n}{12}(2^{2m}-1)
\end{align}
\end{lemma}
\begin{IEEEproof}
Let $\msf{A}$ be a random variable that is uniformly distributed over the fundamental Voronoi cell $\mc{V}_0(\Lambda)$ and independent of $\msf{X}$. Then $\msf{X}+\msf{A}$ is a continuous random variable whose distribution is uniform over a region $\mathcal{R}$ and has zero mean. The average power of $\mathcal{R}$ is
\begin{align}
\E[\|\mathcal{R} \|^2]  &= \E[\|\msf{X}\|^2]+\E[\|\msf{A}\|^2] 
 = \E[\|\msf{X}\|^2]+n\sigma^2(\Lambda),
\end{align}
where the second equality is according to \cite[Eqs (22)-(23)]{1337105}. As $|\Lambda/2^{m}\Lambda| = 2^{nm}$, the region $\mathcal{R}$ consists of $2^{nm}$ numbers of Voronor cells $\mc{V}_{\lambda}(\Lambda)$, where the coset leader is in the cell center.

Let $\msf{D}$ be a random dither that is uniformly distributed over $\mc{V}(2^m\Lambda)$. Then,
\begin{align}\label{eq:app_p_1}
\E[\|[\msf{V} - \msf{D}]\; \text{mod}\;2^m\Lambda\|^2] = \E[\|\msf{D}\|^2]  = n\sigma^2(2^m\Lambda).
\end{align}
Here, $[\msf{V} - \msf{D}]\; \text{mod}\;2^m\Lambda$ becomes a continuous random variable that is uniformly distributed over the fundamental Voronoi cell $\mc{V}_0(2^m\Lambda)$. We note that $\text{Vol}(\mc{V}_0(2^m\Lambda)) = 2^{nm}\text{Vol}(\mc{V}_0(\Lambda))$.

Since $\Lambda = \mathbb{Z}^n$, the lattice partition $\mathbb{Z}^n/2^{m}\mathbb{Z}^n$ is the $n$-fold Cartesian product of the one-dimensional lattice partition $\mathbb{Z}/2^{m}\mathbb{Z}$. The fundamental Voronoi cell $\mc{V}_0(2^m\mathbb{Z}^n)$ is the $n$-fold Cartesian product of the fundamental Voronoi cell $\mc{V}_0(2^m\mathbb{Z})$. Given that the fundamental Voronoi cell $\mc{V}_0(\mathbb{Z})$ is $[-\frac{1}{2},\frac{1}{2}]$, thus the region of $\mc{V}_0(2^m\mathbb{Z})$ is $[-\frac{2^m}{2},\frac{2^m}{2}]$. The coset leaders of $\mathbb{Z}^n/2^m\mathbb{Z}^n$ are the $n$-fold Cartesian product of the one-dimensional coset leaders of $\mathbb{Z}/2^m\mathbb{Z}$, i.e., $\{0,\ldots, 2^{m}-1\}$. After subtracting a fixed dither to ensure zero mean, the one-dimensional coset leaders become $\{-\frac{2^m-1}{2},\ldots, \frac{2^m-1}{2}\}$. We note that the \emph{union} of the Voronoi cells of these coset leaders is exactly $[-\frac{2^m-1}{2}-\frac{1}{2},\frac{2^m-1}{2}+\frac{1}{2}]$, same with the fundamental Voronoi cell $\mc{V}_0(2^m\mathbb{Z})$. Hence, the Cartesian products of both the union cells and the fundamental Voronoi cell $\mc{V}_0(2^m\mathbb{Z})$ lead to the same support.
Thus, the random variable uniformly distributed over these regions have the same average power, meaning that $\E[\|\mathcal{R} \|^2] = \E[\|\msf{D}\|^2]$.

As a result, the average power of $\msf{X}$ is therefore
\begin{align}
\E [\| \msf{X}\|^2] = n(\sigma^2(2^m\Lambda)-\sigma^2(\Lambda)) =& n(2^{2m}-1)\text{Vol}(\Lambda)^{\frac{2}{n}}\psi(\Lambda)) \nonumber \\
=& \frac{n}{12}(2^{2m}-1),
\end{align}
where $\text{Vol}(\Lambda) = 1$ and $\psi(\Lambda) = \frac{1}{12}$ for $\Lambda = \mathbb{Z}^n$.
\end{IEEEproof}

\begin{lemma}\label{lem:dmindp}
For the $n$-dimensional ideal lattice $\Lambda$ constructed via cyclotomic construction, there exists at least a lattice point $\boldsymbol{\lambda} \in \Lambda$ and $\boldsymbol{\lambda} \neq  \mathbf{0} $ satisfying both $d_E(\boldsymbol{\lambda},\mathbf{0}) = d_{E,\min}(\Lambda)$ and $d_{p}(\boldsymbol{\lambda},\mathbf{0}) = d_{p,\min}(\Lambda)$.
\end{lemma}
\begin{IEEEproof}
We consider the lattice point $\boldsymbol{\lambda}$ generated from a length $n$ integer vector $\mathbf{b} = [0,0,\ldots,1]$. 
Since the generator matrix of $\Lambda$, $\mathbf{G}_{\Lambda} $ is a rotated version of $\mathbf{I}_{n}$, i.e., the rotation matrix itself, it is obvious that $d_E(\boldsymbol{\lambda},\mathbf{0}) = d_E(\mathbf{b},\mathbf{0})= d_{E,\min}(\mathbb{Z}^n) = d_{E,\min}(\Lambda)$ because rotation does not affect the Euclidean distance.

Now, we write the analytical expression for $\boldsymbol{\lambda}$ as
\begin{align}
\boldsymbol{\lambda} =& \mathbf{b} \mathbf{G}_{\Lambda} 
 \overset{\eqref{G_ideal}}= [0,0,\ldots,1]\cdot\frac{1}{\sqrt{p}}\mathbf{T}\cdot [(\sigma_j(\zeta^i+\zeta^{-i}))_{i,j=1}^n] \nonumber \\
 &\cdot \text{diag}(\sqrt{\sigma_1(\varsigma)},\ldots,\sqrt{\sigma_n(\varsigma)}) \nonumber \\
 = &[0,0,\ldots,1]\cdot\frac{1}{\sqrt{p}}\cdot [(\sigma_j(\zeta^i+\zeta^{-i}))_{i,j=1}^n] \nonumber \\
&\cdot\text{diag}(\sqrt{\sigma_1(\varsigma)},\ldots,\sqrt{\sigma_n(\varsigma)}) \nonumber \\
 =& \frac{1}{\sqrt{p}}[\sqrt{\sigma_1(\varsigma)}\sigma_1(\zeta^n+\zeta^{-n}),\sqrt{\sigma_2(\varsigma)}\sigma_2(\zeta^n+\zeta^{-n}),\nonumber \\
&\ldots,\sqrt{\sigma_n(\varsigma)}\sigma_n(\zeta^n+\zeta^{-n})].
\end{align}
As $\zeta$ is the $p$-th root of unity $e^{\frac{2\pi\sqrt{-1}}{p}}$ of the polynomial $p(z) = z^p - 1 =\prod_{j=0}^{p-1}(z - e^{\frac{2j\pi\sqrt{-1}}{p}})$, thus
\begin{align}
-p(-z) &= z^p + 1 = \prod\nolimits_{j=1}^{p-1}\left(z+e^{\frac{2j\pi\sqrt{-1}}{p}}\right) \nonumber \\
&= (z+1)\prod\nolimits_{j=1}^{\frac{p-1}{2}}\left[\left(z+e^{\frac{2j\pi\sqrt{-1}}{p}}\right)\left(z+e^{-\frac{2j\pi\sqrt{-1}}{p}}\right)\right].
\end{align}
Substituting $z=1$ into $-p(-z)$ gives
\begin{align}\label{eq:norm1}
1 =& \prod\nolimits_{j=1}^{\frac{p-1}{2}}\left[\left(1+e^{\frac{2j\pi\sqrt{-1}}{p}}\right)\left(1+e^{-\frac{2j\pi\sqrt{-1}}{p}}\right)\right] \nonumber \\
= &\prod\nolimits_{j=1}^{\frac{p-1}{2}}\left(e^{\frac{j\pi\sqrt{-1}}{p}}+e^{-\frac{j\pi\sqrt{-1}}{p}}\right)^2 \nonumber \\
 =& \left(\prod\nolimits_{j=1}^{\frac{p-1}{2}} \left|e^{\frac{j\pi\sqrt{-1}}{p}}+e^{-\frac{j\pi\sqrt{-1}}{p}}\right|\right)^2 \nonumber \\
 \overset{(a)}=& \left(\prod\nolimits_{j=1}^{n} \left|e^{\frac{2jn\pi\sqrt{-1}}{2n+1}}+e^{-\frac{2jn\pi\sqrt{-1}}{2n+1}}\right|\right)^2 \nonumber \\
\Rightarrow\; 1  =& \prod\nolimits_{j=1}^{n} \left|\zeta^{jn}+\zeta^{-jn}\right|,
\end{align}
where $(a)$ is due to shifting the periodic function $|e^{\frac{j\pi\sqrt{-1}}{p}}+e^{-\frac{j\pi\sqrt{-1}}{p}}|$ to the right by $j\pi$ and substituting $p = 2n+1$. The product distance between $\boldsymbol{\lambda}$ and $\mathbf{0}$ is then computed as
\begin{align}
d_{p}(\boldsymbol{\lambda},\mathbf{0})  =& \prod\nolimits_{j=1}^n |\lambda_j| \nonumber \\
=& \left|\left(\frac{1}{\sqrt{p}}\right)^n \sqrt{\prod\nolimits_{j=1}^n \sigma_j(\varsigma)} \prod\nolimits_{j = 1}^n \sigma_j(\zeta^n+\zeta^{-n})\right| \nonumber \\
 =& \left(\frac{1}{\sqrt{p}}\right)^n \sqrt{N(\varsigma)}  \prod\nolimits_{j = 1}^n |(\zeta^{jn}+\zeta^{-jn})| \nonumber \\
 \overset{(a)}=& \left(\frac{1}{\sqrt{p}}\right)^n \sqrt{p}   
 = p^{-\frac{n-1}{2}} \overset{\eqref{eq:dpcal}}=d_{p,\min}(\Lambda),
\end{align}
where $(a)$ follows that $N(\varsigma) = p$ as this is a necessary condition to obtain the $\mathbb{Z}^n$ ideal lattice \cite[Eq. (7.4)]{Oggier:2004:ANT:1166377.1166378} and $\prod\nolimits_{j = 1}^n |(\zeta^{jn}+\zeta^{-jn})| = 1$ follows from \eqref{eq:norm1}.
\end{IEEEproof}

\begin{lemma}\label{the:1a}
Consider any two layers: layer 1 and 2 in $\Lambda$. Let $\mathbf{a}$ and $\mathbf{b}$ be two distinct points on layer 1 and $\mathbf{e}$ and $\mathbf{f}$ be two distinct points on layer 2, such that $d_E(\mathbf{a,b}) = d_E(\mathbf{e,f})$ and $d_p(\mathbf{a,b}) \leq d_p(\mathbf{e,f})$, where $d_E$ and $d_p$ denote the Euclidean distance and product distance, respectively. Then for \emph{any} two distinct points $\mathbf{a}'$ and $\mathbf{b}'$ on layer 1 and \emph{any} two distinct points $\mathbf{e}'$ and $\mathbf{f}'$ on layer 2 satisfying $d_E(\mathbf{a',b'}) = d_E(\mathbf{e',f'})$, the following relation holds:
\begin{align}
d_p(\mathbf{a',b'}) \leq d_p(\mathbf{e',f'}).
\end{align}
\end{lemma}
\begin{IEEEproof}
For layer 1 and 2, we can write the corresponding $n$-dimensional line equations as
$\mathbf{l}_1 = t_1(\mathbf{b}-\mathbf{a})+\mathbf{a}, t_1 \in \mathbb{R}$ and 
$\mathbf{l}_2 = t_2(\mathbf{f}-\mathbf{e})+\mathbf{e},t_2 \in \mathbb{R}$,
respectively. Since points $\mathbf{a}'$ and $\mathbf{b}'$ is in layer 1, we have
$\mathbf{a}' = t_a(\mathbf{b}-\mathbf{a})+\mathbf{a}, t_a \in \mathbb{R}, \; 
\mathbf{b}' = t_b(\mathbf{b}-\mathbf{a})+\mathbf{a},t_b \in \mathbb{R}$.

Similarly, for layer 2 with points $\mathbf{e}'$ and $\mathbf{f}'$ on it, we have
$\mathbf{e}' = t_e(\mathbf{f}-\mathbf{e})+\mathbf{e}, t_e \in \mathbb{R}, \; 
\mathbf{f}' = t_f(\mathbf{f}-\mathbf{e})+\mathbf{e},t_f \in \mathbb{R}$.
The product distance between points $\mathbf{a}$ and $\mathbf{b}'$ is given by
\begin{align}
&d_{p}(\mathbf{a',b'}) = \prod\nolimits_{i=1}^n|a'_i-b'_i| = \prod\nolimits_{i=1}^n|(b_i-a_i)(t_a-t_b)| \nonumber \\
&= |t_a-t_b|^n\prod\nolimits_{i=1}^n|b_i-a_i| 
 = |t_a-t_b|^nd_{p}(\mathbf{a,b}).
\end{align}
Similarly, the product distance point $\mathbf{e}'$ and $\mathbf{f}'$ is given by
\begin{align}
&d_{p}(\mathbf{e',f'}) = \prod\nolimits_{i=1}^n|e'_i-f'_i| = \prod\nolimits_{i=1}^n|(f_i-e_i)(t_e-t_f)| \nonumber \\
&= |t_e-t_f|^n\prod\nolimits_{i=1}^n|f_i-e_i| 
 = |t_e-t_f|^nd_{p}(\mathbf{e,f}).
\end{align}
Since $d_E(\mathbf{a',b'}) = d_E(\mathbf{e',f'})$, we have
\begin{align}\label{eq:dmin_eq}
&\sqrt{\sum\nolimits_{i=1}^n(a'_i-b'_i)^2} =  \sqrt{\sum\nolimits_{i=1}^n(e'_i-f'_i)^2}, \nonumber \\
\Rightarrow& \; \sqrt{\sum\nolimits_{i=1}^n((b_i-a_i)(t_a-t_b))^2} =  \sqrt{\sum\nolimits_{i=1}^n((f_i-e_i)(t_e-t_f))^2} \nonumber \\
\Rightarrow&  \;|t_a-t_b|\sqrt{\sum\nolimits_{i=1}^n(b_i-a_i)^2} =  |t_e-t_f|\sqrt{\sum\nolimits_{i=1}^n(f_i-e_i)^2} \nonumber \\
\overset{(a)}\Rightarrow& \; |t_a-t_b| =  |t_e-t_f|,
\end{align}
where $(a)$ follows that $d_E(\mathbf{a,b}) = d_E(\mathbf{e,f})$. Since $d_p(\mathbf{a,b}) \leq d_p(\mathbf{e,f})$, and based on \eqref{eq:dmin_eq}, we have
$|t_a-t_b|^nd_{p}(\mathbf{a,b}) \leq |t_e-t_f|^nd_{p}(\mathbf{e,f})$,
which implies that $d_p(\mathbf{a',b'}) \leq d_p(\mathbf{e',f'})$.
\end{IEEEproof}

\begin{lemma}\label{lem:1}
Let $\mathbf{a}$, $\mathbf{b}$ and $\mathbf{c}$ be three points on a line in $\mathbb{R}^n$. Assume that point $\mathbf{b}$ is located in between points $\mathbf{a}$ and $\mathbf{c}$. Then, the product distances of line segments $\mathbf{ab}$, $\mathbf{bc}$ and $\mathbf{ac}$ satisfy
\begin{align}
\sqrt[n]{d_{p}(\mathbf{a,c})} = \sqrt[n]{d_{p}(\mathbf{a,b})}+\sqrt[n]{d_{p}(\mathbf{b,c})}.
\end{align}
\end{lemma}
\begin{IEEEproof}
Let $\mathbf{a} = [a_1,\ldots,a_n]$, $\mathbf{b} = [b_1,\ldots,b_n]$ and $\mathbf{c} = [c_1,\ldots,c_n]$. The equation of the $n$-dimensional line through point $\mathbf{a}$ to point $\mathbf{b}$ is
$\mathbf{l} = t(\mathbf{b}-\mathbf{a})+\mathbf{a},t \in \mathbb{R}$.
Here, the direction of the line is from $\mathbf{a}(t=0)$ to $\mathbf{b}(t=1)$. Since point $\mathbf{c}$ is also on this line, thus point $\mathbf{c}$ satisfies
\begin{align}\label{eq:n_line}
\mathbf{c} = t'(\mathbf{b}-\mathbf{a})+\mathbf{a},t'\in \mathbb{R}.
\end{align}
Since point $\mathbf{b}$ is located in between point $\mathbf{a}$ and $\mathbf{c}$, we have $t'>1$ to ensure that the directions from $\mathbf{a}$ to $\mathbf{b}$ and from $\mathbf{b}$ to $\mathbf{c}$ are the same.
The $n$-th square root of the product distance of line segment $\mathbf{ac}$ is given by
\begin{align}\label{eq:p1}
\sqrt[n]{d_{p}(\mathbf{a,c})} = \sqrt[n]{\prod\nolimits_{i=1}^n|a_i-c_i|}
\end{align}
The $n$-th square roots of product distances of line segment $\mathbf{ab}$ is
\begin{align}\label{eq:p2}
\sqrt[n]{d_{p}(\mathbf{a,b})} &= \sqrt[n]{\prod\nolimits_{i=1}^n|a_i-b_i|} \overset{\eqref{eq:n_line}}= \sqrt[n]{\prod\nolimits_{i=1}^n|\frac{a_i-c_i}{t'}|} \nonumber \\
&= \frac{1}{t'}\sqrt[n]{\prod\nolimits_{i=1}^n|a_i-c_i|}
\end{align}
We also note that the $n$-th square roots of product distances of line segment $\mathbf{bc}$ is
\begin{align}\label{eq:p3}
\sqrt[n]{d_{p}(\mathbf{b,c})} &= \sqrt[n]{\prod\nolimits_{i=1}^n|b_i-c_i|} \overset{\eqref{eq:n_line}}= \sqrt[n]{\prod\nolimits_{i=1}^n|\frac{c_i-a_i}{t'}+a_i-c_i|} \nonumber \\
&= |\frac{1}{t'}-1|\sqrt[n]{\prod\nolimits_{i=1}^n|a_i-c_i|} \nonumber \\
&= (1-\frac{1}{t'})\sqrt[n]{\prod\nolimits_{i=1}^n|a_i-c_i|}
\end{align}
Noting that \eqref{eq:p2}$+$\eqref{eq:p3}$=$\eqref{eq:p1} completes the proof.
\end{IEEEproof}

\bibliographystyle{IEEEtran}
\bibliography{MinQiu}

\end{document}